\documentclass[fleqn,usenatbib]{mnras}

\usepackage{newtxtext,newtxmath}

\usepackage[T1]{fontenc}
\usepackage{ae,aecompl}

\usepackage{graphicx}	
\usepackage{amsmath}	

\title[Photohadronic modelling of Mrk 421]{Photohadronic Modelling of the 2010 Gamma-ray Flare from Mrk~421}

\author[A. Rosales de Le\'{o}n et al.]{Alberto Rosales de Le\'{o}n\thanks{Contact e-mail: alberto.rosales-de-leon@durham.ac.uk},
Anthony M. Brown
and Paula M. Chadwick
\\
$^{1}$Centre for Advanced Instrumentation (CfAI), Department of Physics, University of Durham, South Road, Durham DH1 3LE, UK. }

\date{Published 2020 December 12. Accepted 2020 December 9. Received 2020 December 1; in original form 2020 August 3}

\pubyear{2020}

\begin{document}
\label{firstpage}
\pagerange{\pageref{firstpage}--\pageref{lastpage}}
\maketitle

\begin{abstract}
Blazars are a subclass of active galactic nuclei (AGN) that have a relativistic jet with a small viewing angle towards the observer. Recent results based on hadronic scenarios have motivated an ongoing discussion of how a blazar can produce high energy neutrinos during a flaring state and which scenario can successfully describe the observed gamma-ray behaviour. Markarian 421 is one of the closest and brightest objects in the extragalactic gamma-ray sky and showed flaring activity over a 14-day period in March 2010. In this work, we describe the performed analysis of \textit{Fermi}-LAT data from the source focused on the MeV range (100 MeV - 1 GeV), and study the possibility of a contribution coming from the $p\gamma$ interactions between protons and MeV SSC target photons to fit the very high energy (VHE) gamma-ray emission. The fit results were compared with two leptonic models (one-zone and two-zone) using the Akaike Information Criteria (AIC) test, which evaluates goodness-of-fit alongside the simplicity of the model. In all cases the photohadronic model was favoured as a better fit description in comparison to the one-zone leptonic model, and with respect to the two-zone model in the majority of cases. Our results show the potential of a photohadronic contribution to a lepto-hadronic origin of gamma-ray flux of blazars. Future gamma-ray observations above tens of TeV and below 100 MeV in energy will be crucial to test and discriminate between models.
\end{abstract}



\section{Introduction}
\label{sec:introduction}

 Blazars are a sub-class of radio-loud Active Galactic Nuclei (AGN) with a relativistic jet pointing close to our line of sight. These objects have a highly variable spectrum and flaring states, periods of enhanced activity on time scales that can go from hours to months. This variability makes it extremely difficult to model the broad-band spectral energy distribution (SED) of these sources.

Leptonic models have been used for many years to fit the SEDs from blazars (see, for example, \cite{Bloom:1996,Tavecchio:1998,Sahayan:2011,Aleksic:2014rca}). These models consider a population of relativistic electrons to be responsible for the characteristic two peaked SED of a blazar. In this approach, the first peak (covering radio to X-rays) can be explained by synchrotron emission; meanwhile, the second peak (X-rays and gamma rays) may be composed of different contributions coming from inverse Compton interactions between the electrons and a photon field. One option is  synchrotron-self Compton contributions (SSC) coming from the emission region inside the jet \citep{Maraschi:1993, Dermer:1993, Sikora:1994, Bloom:1996, Tavecchio:1998}; another option is to consider external Compton-scattering contributions with the target photons coming from the broad-line region (BLR), the accretion disk, the dusty torus or any other external radiation field \citep{Boettcher:2013wxa,Finke:2016vts}.

The different versions of the leptonic scenarios mentioned above have been applied to the spectra of several objects, but the challenging observations raise the question of whether a hadronic component is necessary to explain the full spectrum. This idea has once again caught the interest of the scientific community after a possible correlation between the blazar TXS~0506+056 and a neutrino alert (IC-170922A). On September 22nd 2017, an extensive multi-wavelength campaign was triggered by the high energy neutrino-induced muon track event. The blazar TXS~0506+056 was reported to be $0.1^{\circ}$ from the best-fitting neutrino direction by the \textit{Fermi} Large Area Telescope (\textit{Fermi}-LAT) Collaboration \citep{MMS-Tanaka-Tel}. This source was in a flaring state at the time and had a considerably brightened in the GeV band since April 2017. After the follow-up observations, models associating neutrino and gamma-ray production during the flaring state of the source found the gamma-ray emission was correlated with the neutrino alert at a statistical significance of $3 \sigma$ \citep{MultiW:2018dnn}. In addition, the IceCube Collaboration performed an independent analysis using prior data. The result was an excess of high energy neutrino events coming from the direction of the source, with respect to atmospheric backgrounds, at a significance level of $\sim 3.5\sigma$ between September 2014 and March 2015 \citep{IceCube:2018cha}. These findings motivated an ongoing discussion of how a blazar can reproduce the experimental data and which model (or models) can successfully describe the observed behaviour (e.g. \cite{Righi:2018,Murase:2018iyl,Palladino:2019,Padovani:2019xcv,Rodrigues:2018tku,Cerruti:2018tmc,Petropoulou:2019zqp,Winter:2019hee,Ruo:2019,Halzen:2018iak}).

The possible neutrino/gamma-ray connection exhibited between the IC-170922A alert and TXS 0506+056 is one motivation to explore hadronic contributions, but there are other hints and plausible evidence for this scenario. The blazar 3HSP J095507.9+355101 was recently reported to be in a flaring state and $0.62^{\circ}$ away from the best-fit position of the alert IceCube200107A, a muon track event \citep{2020arXiv200306405G}. There is also PKS B1424-418, a Flat Spectrum Radio Quasar (FSRQ) which was in temporal and positional coincidence with a high-energy starting event (HESE) on December 2012, a cascade-like event with a reconstructed energy of 2 PeV but a median positional uncertainty of $\sim 16^{\circ}$ that gives an estimated $\sim 5 \%$ chance of coincidence \citep{Kadler:2016ygj}. In addition, there are several studies searching for gamma-ray counterparts and predicting consistent limits within the IC neutrino flux so far \citep{Krauss:2014tna, Krauss:2015pja, Brown:2015zxa, Padovani:2016wwn, Glusenkamp:2015jca}; the successful SED modelling of blazars during flaring episodes using hadronic models \citep{Mucke:2000rn, Diltz:2015kha, Diltz:2016aqg, Sahu:2016bdu, Sahu:2018rqs, Sahu:2018gik}; and hadronic emission has been proposed as an explanation for the spectral hardening in TeV energy gamma-ray spectra, behaviour which has been observed in some blazars (e.g. W~Comae, 3C~66A \citep{Boettcher:2013wxa}, 1ES~0229+200 \citep{Tavecchio:2009}, 1ES~1101-232 and H~2356-309 \citep{Aharonian:2005gh}). 

Hadronic scenarios propose that protons are accelerated to relativistic energies in blazar jets. A group of models invoke photohadronic ($p\gamma$) interactions, which involve collisions between the high energy protons and a target photon field. Another option is hadronuclear interactions ($p p$), where a matter target such as a gas cloud is required. Both scenarios lead to photo-meson production, from which gamma rays and neutrinos are generated in the decay process. The decay products will also emit radiation, including proton-synchrotron emission, photo-pion production, electron-positron synchrotron triggered pair cascades, or even synchrotron from the charged decay products (muons and pions). Depending on the hadronic model, the physical conditions and chosen parameters, there might be a dominant component, for instance proton-synchrotron radiation \citep{Muecke:2002bi, Dimitrakoudis:2012} or photo-pion production \citep{Mannheim:1992, Mannheim:1993}.

In the hadronuclear models, $pp$ interactions can occur, if the high energy protons accelerated along the jet reach a matter target, for example the gas clouds in the broad line region (BLR) around the AGN \citep{Dar:1996qv, Araudo:2010, Ruo:2019}. This process has a lower interaction cross section compared to the prominent $\Delta-$resonance of the photohadronic process, so a high density target is required to improve the efficiency of the hadronic interactions.

If a model considers mixed contributions from hadronic and leptonic origin, then it can also be referred as lepto-hadronic (including some of the previously given examples: \cite{Araudo:2010, Diltz:2016aqg, Rodrigues:2018tku, Cerruti:2018tmc, Petropoulou:2019zqp, Ruo:2019}; among others).

In this work, we test the potential of a dominant photohadronic contribution within a lepto-hadronic scenario to fit the very high energy (VHE) gamma-ray observations (E > 100~GeV) during a flaring blazar state. We account for the $p\gamma$ interactions through the $\Delta-$resonance approximation; the subsequent decay products include gamma rays and neutrinos in the following way:

\begin{equation}
p + \gamma \rightarrow \Delta^{+} \rightarrow \bigg\{ \begin{array}{c}
p \pi^{0}, \; \pi^{0} \rightarrow \gamma \gamma \\
n \pi^{+}, \; \pi^{+} \rightarrow \mu^{+} \nu_{\mu}, \; \mu^{+} \rightarrow e^{+} + \nu_{e} + \bar{\nu_{\mu}} \, . \\ \end{array} 
\label{Eq-1}
\end{equation}

The $\Delta^{+}$ particle decays into $(p + \pi^0)$ in 2/3 of all cases while goes to $(n + \pi^{+})$ in 1/3 of all cases \citep{Hummer:2010vx}. The model considered is described in Section~\ref{sec:model} and a broader discussion of the $\Delta-$resonance can be found in \cite{Mucke:1998mk, Gaisser:1994yf}.

The blazar Markarian 421 (Mrk 421; RA=66.114$^{\circ}$, Dec=38.209$^{\circ}$, z=0.031) is one of the closest and brightest objects in the extragalactic VHE sky. It was the first extragalactic source detected using Imaging Atmospheric Cherenkov Telescopes (IACTs) \citep{Punch:1992} and has been regularly monitored since then. Mrk~421 has been measured during flaring states on several occasions (e.g. \cite{Blaz2005,Abdo2011ApJ,Aleksic:2015,Aleksic:2014rca}), and the recorded multi-wavelength (MWL) data from radio to high energy gamma rays makes it an ideal candidate to test different production mechanisms and their evolution during a flare. The 2010 flaring activity from Mrk~421 \citep{Aleksic:2014rca} provides a rich dataset. We performed an analysis in the MeV energy range (100~MeV-1~GeV) with the updated instrument response functions (IRFs) from the \textit{Fermi}-LAT to obtain an input seed photon spectrum to a photohadronic model that could provide a good fit to the VHE gamma-ray data.

This paper is structured as follows: in Section~\ref{sec:flare} we review the 2010 flare, in Section~\ref{sec:analysis} we describe the \textit{Fermi}-LAT analysis undertaken, and in Section~\ref{sec:fermi_results} we describe the results of that analysis. Section~\ref{sec:model} describes the photohadronic model and the method used to fit this to the data, and Section~\ref{sec:model_results} describes the results of this fit. Finally, we discuss our results in Section~\ref{sec:discussion} and conclude in Section~\ref{sec:conclusions}. 

\section{Flaring Activity in 2010}
\label{sec:flare}
Mrk 421 exhibited flaring activity over a 14-day period in 2010 from March 10 to March 22 (MJD~55264-55277). At the time, a multi-instrument campaign was performed which included the gamma-ray space telescope \textit{Fermi}-LAT and three IACTs: the Major Atmospheric Gamma-ray Imaging Cherenkov (MAGIC) telescope system, the Very Energetic Radiation Imaging Telescope Array System (VERITAS) and the Whipple gamma-ray telescope.

MAGIC took 11 observations in stereoscopic mode with exposure times ranging from 10 to 80 min each, which led to 4.7 h of good-quality data with a zenith angle range of $5^{\circ}-30^{\circ}$. The data collected were taken in dark conditions and were not affected by moonlight, but the data recorded on MJD~55272 and 55275 suffered from bad weather and were therefore removed from the MWL observations \citep{Aleksic:2014rca}. For more details on the MAGIC telescope system see \cite{Aleksic:2012}. 

VERITAS monitored the source on MJD~55260, 55265, and 55267-55274 with a 10 min run per day. The observations were performed at zenith angles $18^{\circ}-23^{\circ}$ to benefit from the lowest possible energy threshold. Further information about the VERITAS instrument can be found in \cite{Perkins2009}.

The Whipple telescope performed 10 observations in ON/OFF and TRK (tracking) modes \citep{Pichel:2009}, lasting from one to six hours each on MJD~55267-55271 and MJD~55273-55277. The dataset collected for this flaring period amounts to 36~h. More information and details about the Whipple telescope can be found in \cite{Kildea2007}. 

\begin{figure}
    \centering
    \resizebox{1.0 \columnwidth}{!}{
        \includegraphics{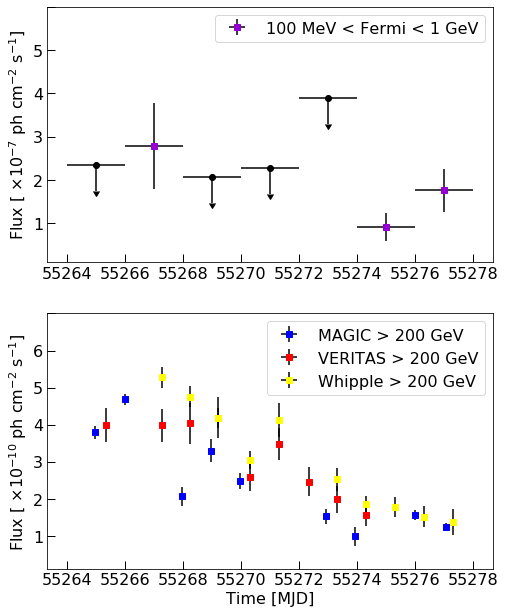}}
    \caption{Light curve of Markarian 421 during the 14 day flaring period in March 2010. The upper plot shows the gamma-ray flux in the energy range 100~MeV < $E{\gamma}$ < 1~GeV  and the points are calculated in couple of days bins. In the bins with a TS<25 upper limits for the flux are shown. The lower plot presents the light curves for MAGIC, VERITAS and Whipple above 200~GeV.}
    \label{LightCurves}
\end{figure}

The VHE gamma-ray data from ground based IACTs published in \cite{Aleksic:2014rca} were used to test a possible dominant contribution from photohadronic interactions. In this paper, we focus on the modelling of the VHE gamma-ray observations. In this case, the \textit{Fermi} analysis and the IACT data were the two key elements needed, and data from other wavelengths were not critical for our calculations (see \cite{Aleksic:2014rca} for a full description of the MWL observations). The light curves from MAGIC, VERITAS and Whipple above 200~GeV are shown in Figure \ref{LightCurves}. In terms of simultaneity of the data, the shorter observation times of the IACTs are embedded in the 2-day bins period defined for the \textit{Fermi} analysis, noting that there is a 7-h time difference between the VERITAS/Whipple and MAGIC observations due to their different longitudes. The variability reported in the gamma-ray data corresponds to daily changes in the VHE emission; no intra-night variability was reported on the days studied.

The photohadronic model does not aim to describe the whole SED with purely hadronic components. This approach relies on a standard leptonic scenario to explain the low energy peak of the SED and provide seed photons for the $p\gamma$ interactions. To describe the blob from which the gamma-ray photons are produced, we have adopted the values of the physical parameters in the one-zone model of \cite{Aleksic:2014rca} relating to magnetic field, Doppler factor and the radius of the emission region. These parameters are fixed during the flaring events, and the evolution of the spectral parameters provides snapshots on the different days considered for modelling.

Mrk 421 was highly active during other months in 2010, and the VERITAS Collaboration reported another flare in February 2010 (MJD~55234-55240), the brightest ever observed from this object in VHE gamma rays \citep{Abeysekara:2020}. They concluded that the time variability of the source is difficult to explain using a single-zone SSC model. This result provides another motivation to try to extend the current models and look for hadronic contributions.

\section{\textit{Fermi} Analysis}
\label{sec:analysis}
Launched in June 2008, the Large Area Telescope (LAT) on board the \textit{Fermi} satellite is a pair conversion telescope covering the energy range from 20 MeV to more than 300 GeV. The LAT's combination of a wide field of view (FoV $\simeq 2.4$~sr) and large effective area allows it to scan the entire gamma-ray sky approximately every 3 hours \citep{Atwood:2009}. The data analysis described here was performed with the Pass8v6 version of the IRF and the v11r5p3 Science Tools software with \textit{Fermipy} \citep{Wood:Fermipy:2017}. This IRF provides a full reprocessing of the entire mission dataset, including improved event reconstruction, a wider energy range, better energy measurements, and significantly increased effective area in comparison to previous versions. As a result, we were able to extend our analysis into a lower energy range than previously possible, resulting in a more comprehensive spectrum of the source during the flaring period.

The \textit{Fermipy} package provides a set of tools and an interface (GTAnalysis) to perform the data preparation, modelling, statistics and analysis tasks. The 4FGL-DR2 catalog \citep{4FGL:2019}, containing the positions and spectral information of the known gamma-ray sources, was used during the analysis.

Our data reduction steps considered all `SOURCE'\footnote{See \url{https://fermi.gsfc.nasa.gov/ssc/data/analysis/documentation/Cicerone/Cicerone_Data/LAT_DP.html}} class events photons in an energy range of 100~MeV to 1~GeV between 2010 March 9th to 22nd (MJD~55264-55277). The flaring period studied was divided into shorter 2-day intervals; this time period was the minimum to obtain enough photon events to calculate the SED points and upper limits. The \textit{Fermi}-LAT data are publicly-available and were downloaded from the LAT data server system\footnote{See \url{https://fermi.gsfc.nasa.gov/cgi-bin/ssc/LAT/LATDataQuery.cgi}}.

The analysis was performed inside a 15$^{\circ}$ region of interest (RoI) around Mrk 421's coordinates, a 90$^{\circ}$ zenith cut angle was applied to avoid gamma rays coming from Earth's atmosphere. The gamma-ray data were then binned using 5 bins per decade in energy and 0.1$^{\circ}$ spatial bin size. To remove sub-optimal data, only the events within good time intervals (GTI) were analysed, these were selected by the `gtmktime' tool filters `(DATA\_QUAL>0) \&\& (LAT\_CONFIG==1)' and a cut above 52$^{\circ}$ in rocking angle.

A model consisting of gamma-ray point sources and a background with a Galactic diffuse and an extragalactic component was employed. All the sources listed in the 4FGL catalog inside a 20$^{\circ}$ neighbourhood from the centre of the ROI were included; this was to account for the possible contributions of sources near the edge of our ROI. The spatial model, position and spectral parameters of the sources were adopted from the 4FGL catalog. The Galactic diffuse emission component used in the model was `gll\_iem\_v06.fits'\footnote{\label{note-2} See \url{https://fermi.gsfc.nasa.gov/ssc/data/access/lat/BackgroundModels.html}}. The extragalactic gamma-ray contribution coming from unresolved extragalactic sources, and residual (misclassified) cosmic-ray emission was also included as an isotropic spectral template parametrised in the file `iso\_P8R2\_SOURCE\_V6\_v06.txt'\textsuperscript{\ref{note-2}}. In the 4FGL catalog, Mrk 421 (4FGL J1104.4+3812) is listed as a source with a log-parabola (LP) spectrum type. However, over the short time intervals which we are considering, there were not sufficient photon statistics to allow a log-parabola model to be distinguished from a power-law (PL) model, so the latter was adopted. The spectral models should not differ significantly in the range of interest for our analysis and the SED extrapolation below 100~MeV.

A maximum likelihood method\footnote{The maximum-likelihood test statistics (TS) is defined as $TS=2[logL-logL_0]$ where $L$ and $L_0$ are the likelihood when the source is included or not, respectively \cite{Mattox:1996}.} was used for fitting the ROI; the spectral shape parameters of the sources were left free to vary within a 5$^{\circ}$ radius around the ROI's centre. The two background components were also left free to vary during the maximum likelihood fitting. The sources in the model that were considered insignificant (TS < 1) were discarded. A second optimisation and fit steps were applied to the ROI using the new model with the same free components and criteria used before.

To check for any point sources inside the ROI which are not listed in the 4FGL catalog, the `find\_sources' routine built within \textit{Fermipy} was implemented. No significant extra source candidates (TS>16) were found during the analysis. 

The SED points for each 2-day bin were calculated using the `sed()' tool included in \textit{Fermipy}, which performs an independent maximum likelihood fit per energy bin for the flux normalisation. In our analysis, the energy range studied (100 MeV to 1 GeV) was split into 5 evenly spaced log-energy bins. The same analysis steps described above were applied to each 2-day interval in the flaring period studied (MJD~55264-55277) to obtain the corresponding SED and PL spectral parameters.

\section{\textit{Fermi} Results}
\label{sec:fermi_results}
\begin{table*}
\centering
\begin{tabular}{cccccc}
\hline
Time & TS & Flux & $N_{\rm PL}$& $\kappa$ \\
MJD & & [$10^{-7}$ ph cm$^{-2}$ s$^{-1}$]& [$10^{-11}$ MeV cm$^{-2}$ s$^{-1}$]& \\
\hline
55266-67& 29 & 2.78 $\pm$ 0.98 & 2.11 $\pm$ 1.32 & 2.21 $\pm$ 0.44 \\
55274-75& 42& 0.91 $\pm$ 0.33 & 4.51 $\pm$ 2.60 & 1.17 $\pm$ 0.61 \\
55276-77& 47& 1.76 $\pm$ 0.50 & 5.94 $\pm$ 3.12 & 1.11 $\pm$ 0.61 \\
\hline
55264-65& 10 & < 2.34  & - & - \\
55268-69&  6 & < 2.07 & - & - \\
55270-71& 18 & < 2.27 & - & - \\
55272-73& 19 & < 3.89 & - & - \\
\hline
\end{tabular}
\caption{Summary table of the spectral parameters of Mrk 421. The 2-day bins with significant TS values are listed on the top of the table. The 4th and 5th columns correspond to the optimised parameters for a PL fit coming from the \textit{Fermi} analysis performed. The days with a low TS value are listed in the bottom of the table together with the upper limits for the gamma-ray flux.}
\label{Table-1}
\end{table*}

The \textit{Fermi} analysis was done with the purpose of characterising the seed photon spectrum of the source using an extrapolation below the 100~MeV energy range. In the assumed scenario, the seed photons for the $p\gamma$ interactions are expected to be between 2 and 168 MeV (see Section \ref{sec:p-gammafit}). For the \textit{Fermi} analysis, we selected a photon energy range from 100~MeV (the recommended starting energy) and up to 1~GeV, then extrapolate to the lower energy range of interest. Using the updated version of the IRFs, the shortest time bins which allowed the spectrum to be obtained were 2-days in length.

\begin{figure}
    \centering
    \resizebox{1.0 \columnwidth}{!}{
        \includegraphics{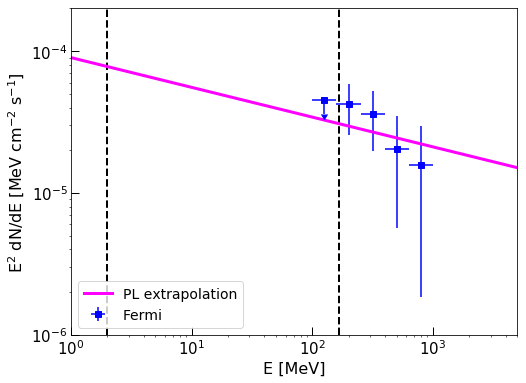}}\\
        (a) MJD 55266-67 \\
    \resizebox{1.0 \columnwidth}{!}{
        \includegraphics{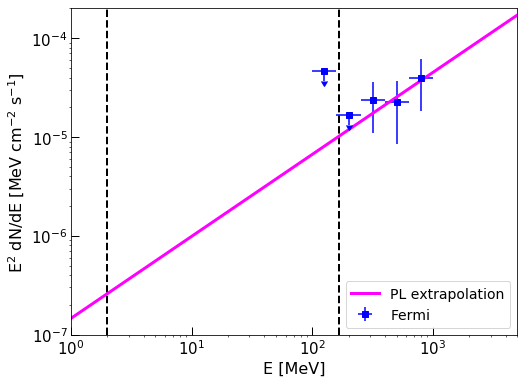}}\\
        (b)  MJD 55274-75  \\
    \resizebox{1.0 \columnwidth}{!}{
        \includegraphics{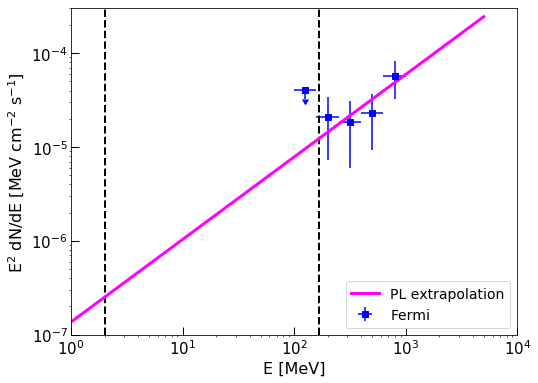}} \\
        (c)  MJD 55276-77 \\

    \caption{\textit{Fermi} spectra (blue points) and power-law extrapolation (magenta line) for the MeV range in 2-days bins: (a) MJD~55266-67, (b) MJD~55274-75, (c) MJD~55276-77. The black dotted vertical lines are positioned at 2 and 168 MeV, which is the expected energy range for the seed photons. The spectral parameters for the selected days are summarised in Table \ref{Table-1}.}
    \label{Fermi-PL}
\end{figure}

As discussed in the previous section, the spectrum in the MeV range was characterised using a PL model: 
\begin{equation}
 \frac{\rm{d} \it{N}}{\rm{d} \epsilon_\gamma} = N_{\rm PL} \epsilon_\gamma^{-\kappa}    ,
\end{equation}
where the normalisation constant $N_{\rm PL}$ and the spectral index $\kappa$ act as free parameters that were optimised to get the best-fit values. The PL description was then used to extend the spectrum below 100 MeV. The spectra of the source alongside the PL extrapolation are shown in Figure \ref{Fermi-PL} for the selected bins.

The spectral parameters of the days on which the source was detected significantly (TS>25) are shown in Table \ref{Table-1} and the corresponding light curves from our \textit{Fermi} analysis and in the VHE energy range are shown in Figure \ref{LightCurves}. This analysis extends the previous results by \cite{Aleksic:2014rca}, which started at 300~MeV rather than 100~MeV. There are no significant flux changes in the VHE band in the combined 2-day bins which we used for the \textit{Fermi} analysis and subsequent modelling. The remaining days presented low photon statistics and were not considered for further VHE fitting with the photohadronic model. In order to get our final result, the gamma-ray spectrum of the source was studied in the MeV energy range, with the fitted spectrum serving as an input for the photohadronic modelling. The time bins in our analysis and that of \cite{Aleksic:2014rca} coincide; however, the extended spectrum analysed with the updated IRF allowed us to calculate a PL extrapolation in our range of interest.

Once the MeV region is characterised, the input seed photon spectrum for the photohadronic model can be expressed as
\begin{equation}
 \Phi_{\rm input} = \epsilon_\gamma^2 \frac{\rm{d} \it{N}}{\rm{d} \epsilon_\gamma}= N_{\rm PL} \epsilon_\gamma^{-\kappa+2}   . 
\end{equation}
 The uncertainty of the seed photon spectrum will impact on the optimisation process of the other free parameters within the model ($\alpha$ and $A_{\gamma}$, see section \ref{sec:p-gammafit}) and therefore the final fitting result.

Alongside the spectral parameters, the light curve of Mrk~421 during the flaring period (MJD~55264-55277) was calculated using the photons in the MeV energy range (100~MeV to 1~GeV), also using 2-day temporal bins (see top of Figure \ref{LightCurves}). The data reduction steps described in the previous section were followed to prepare, optimise and fit the ROI. 

The light curve was generated using the `gta.lightcurve' method within \textit{Fermipy}, which performs a likelihood fit for each time bin. An optimised region of 15 degrees was considered in the energy range of 100~MeV to 1~GeV with 5 bins per decade in energy. The average photon flux for Mrk~421 was $1.67 \pm 0.28 \times 10^{-7}$ ph cm$^{-2}$ s$^{-1}$. For each significant bin (TS>25), the gamma-ray flux is reported in the third column of Table \ref{Table-1}, in bins where the TS value was below 25 we present upper limits.

\section{Photohadronic Contributions to the Flares}
\label{sec:model}
In the lepto-hadronic scenario we consider, a one-zone leptonic model is assumed to contribute to the SED via electron synchrotron and SSC radiation; this is a standard leptonic interpretation. The low-energy peak from the SED comes from the synchrotron component of leptonic origin, while the SSC component is assumed to provide the target photon field in the MeV range. For the photohadronic contribution to arise, it is hypothesised that protons are accelerated into the single spherical emission region of radius $R'_{\rm f}$ (flaring blob), with a tangled magnetic field $B$, propagating along the jet with a velocity $\beta_{\Gamma} c$ and an associated bulk Lorentz factor $\Gamma$. The jet forms a small angle $\theta$ with respect to the line of sight, which results in a Doppler boosting characterised by the Doppler factor ${\cal D}$. It is proposed that during the flaring episode, the blazar possesses a dense, compact inner jet structure \citep{Ghisellini:2005,Marscher:2008, Marscher:2010,MacDonald:2015, MOJAVE:2015, Walker:2016}. Geometrically this represents a double conical shape, with a compact and smaller region enclosed by the jet along its axis (for a schematic view see Figure~1 in \cite{Sahu:2015tua}). The inner compact region has a photon density $n'_{\gamma, \rm{f}}$, which is much higher than the outer region $n'_{\gamma}$, this helps to increase the efficiency of the photohadronic interactions. The prime notation is adopted to refer the jet comoving reference frame.Inside the emission region an electron population will produce synchrotron and SSC radiation following the usual one-zone leptonic scenario. $p\gamma$ interactions can emerge from the collisions between high energy protons and the internal photon field, the SSC photons in the MeV range will serve as targets for the interaction to get a TeV energy gamma rays from the photo-pion production. A more comprehensive review of the photohadronic flaring model can be found in \cite{Sahu:2012wv, Sahu:2013ixa}. 

In this framework, the $\Delta-$resonance approximation is used. The $\Delta^{+}$ particle has a mass of $m_{\Delta}=1.232$ GeV; this is the threshold for interaction and corresponds to the production of the particle at rest. Above this energy threshold, the cross-section of the process is enhanced and this decay channel becomes dominant over other components. At its peak, the cross-section of the $\Delta-$resonance reaches a value of $\sigma_{\rm{peak}} \sim 500$ $\mu\text{barn}$, which is $\sim 5 \times 10^{-28} \text{cm}^2$, being this bigger by a factor of $\sim 5$ than the direct channel cross section production \citep{Hummer:2010vx}.

The threshold of the interaction dictates an energy relation between the proton energy $E'_{\rm p}$ and seed photon $\epsilon'_\gamma$ in the emission region reference frame:
\begin{equation}
    E'_{\rm p} = \frac{m^2_{\Delta} - m^2_{\rm p}}{2 \epsilon'_\gamma (1-\rm{cos}\phi)}  \, ,
\end{equation}

where $m_{\rm p}$ is the proton mass and $\phi$ is the angle formed between the interacting particles. Since the proton will collide with the target photons from all directions there is not a preferred angle of interaction and $1-\rm{Cos}(\phi) \sim 1$. In the observer's frame, due to the Doppler boosting effect from the jet, the proton energy will be enhanced as:
\begin{equation}
    E_{\rm p} = \frac{\Gamma}{1+z} E'_{\rm p} \, ,
\end{equation}
where $E_{\rm p}$ is the energy which would be measured by the observer if the proton could be able to escape the source and reach Earth without energy loss. In a similar way, the target photon energy in the observer's frame can be expressed as:
\begin{equation}
    \epsilon_{\gamma} = \frac{D}{1+z} \epsilon'_{\gamma}  \, .
\end{equation}

Considering that each pion carries $ \sim 20 \%$ of the proton energy \citep{Hummer:2010vx} and in the photo-pion production 2 gamma rays are produced from the $\pi^0$-decay, we have the following relation between the gamma-ray photon energy $E_\gamma$ produced with a proton energy $E_{\rm p}$ in the observer's frame:
\begin{equation}
    E_\gamma = \frac{1}{10} \frac{D}{1+z} E'_{\rm p} = \frac{D}{10 \Gamma} E_{\rm p}  \, .
\label{Eq-7}
\end{equation}

\subsection{Photohadronic Fit}
\label{sec:p-gammafit}
In this scenario, high energy protons can be injected into a confined region (a spherical blob) of radius $R'_{\rm f}$ inside the blazar's jet. For simplicity, it is assumed that the internal jet region and the external jet are moving with almost the same bulk Lorentz factor $\Gamma$. For blazars, we consider that the Doppler factor and the Lorentz factor are approximately of the same magnitude ${\cal D} \sim \Gamma$ \citep{2019:Oikonomou}.

From the energy threshold condition to produce the $\Delta-$resonance, an energy relation between the target photons $\epsilon_{\gamma}$ and the gamma-ray photons $E_{\gamma}$ in the observer's frame can be expressed as:
\begin{equation}
E_{\gamma} \epsilon_{\gamma} \simeq \text{0.032} \frac{{\cal D}^2}{(1+z)^2} \; \text{GeV}^2 \, .
\label{Eegamma}
\end{equation}

The central region of an AGN possesses shocks that are able to accelerate electrons and ions trough the \textit{Fermi} mechanism; when one of these relativistic particles crosses the shock from downstream to upstream or vice versa, it gains energy \citep{FermiShocks}. A PL injected spectrum for the protons is considered: ${\rm d} N (E_{\rm p})/ {\rm d} E_{\rm p} \propto E_{\rm p}^{-\alpha}$, where the spectral index $\alpha $ is treated as a free parameter in the model.
The high energy protons will interact in the inner jet region where the seed photon density is $n'_{\gamma, \rm f}$. The gamma-ray spectrum obtained at VHE will depend proportionally on the photon background and the injected proton spectrum \citep{Sahu:2012wv, Sahu:2013ixa}:
\begin{equation}
    F_{\rm int} (E_{\gamma}) \propto n'_{\gamma, \rm f}  E^2_{\rm p} \frac{\rm{d} \it{N}_{\rm p}}{\rm{d} \it{E}_{\rm p}}  \, .
\end{equation}
The seed photon density will impact on the efficiency of the $p\gamma$ process; a low value reduces the chances of interaction and therefore the gamma-ray photon emission obtained by this method. This photon density in the inner region of the jet is unknown, but we can set a very rough upper limit by assuming that the Eddington luminosity ($L_{\rm Edd}$) of the source should not be exceeded and that it is equally shared by the jet and the counter jet. The upper limit on the seed photon density can be placed using: 
\begin{equation}
    n'_{\gamma, {\rm f}} \ll \frac{L_{\rm Edd}}{8 \pi R'^2_{\rm f} \epsilon'_{\gamma}}  \, ,
\end{equation}
where $L_{\rm Edd}$ for Mrk 421 is $\sim 2.5 \times 10^{46}$ erg s$^{-1}$ for a black hole mass of $2 \times 10^ 8 M_{\odot }$ \citep{Sahu:2015tua}. This gives us a limit of $\sim 3\times 10^{16}$ ph cm$^{-3}$. The photon density $n'_{\gamma, \rm f}$ is proportional to the luminosity $L_\gamma(\epsilon_\gamma)$, and inversely proportional to the seed photon energy $\epsilon_{\gamma}$. The luminosity at a certain energy is proportional to the observed flux $\Phi_{\rm input}(\epsilon_\gamma)$, which is known from the PL extrapolation obtained in Section \ref{sec:fermi_results}, so we have that:
\begin{equation}
n'_{\gamma, \rm f} \propto \Phi_{\rm input} (\epsilon_{\gamma}) \epsilon_{\gamma}^{-1} \, .
\end{equation}
This means the intrinsic gamma-ray flux will follow:
\begin{equation}
    F_{\rm int}(E_{\gamma}) \propto \Phi_{\rm input} (\epsilon_{\gamma}) \epsilon_{\gamma}^{-1} E_{\rm p}^{2} \frac{\rm{d} \it{N}_{\rm p}}{\rm{d} \it{E}_{\rm p}} \, ,
\end{equation} 
considering the PL injected spectrum of the protons and using the energy relations between the proton energy ($E_{\rm p}$), the seed photon energy ($\epsilon_{\gamma}$) and the energy of the gamma-ray photon ($E_{\gamma}$), the intrinsic gamma-ray flux $F_{\rm int}$ coming from the $\pi^0$-decay can be expressed as:
\begin{equation}
    F_{\rm int}(E_{\gamma})=A_{\gamma} \Phi_{\rm input}(\epsilon_{\gamma} )\left (
  \frac{E_{\gamma}}{\text{TeV}}  \right )^{-\alpha+3} \, ,
\label{Int_Flux}
\end{equation}
where $A_{\gamma}$ is a dimensionless normalisation constant that absorbs the information from the various proportional relations given above, and $\alpha$ is the power index from the assumed proton spectrum. In this methodology, $A_{\gamma}$ and $\alpha$ are optimised to fit the VHE gamma-ray data day by day.

When calculating the gamma-ray spectra, we must account for the attenuation of the high energy gamma rays due to the pair production effect with the Extragalactic Background Light (EBL). The EBL provides an attenuation factor of the form $e^{-\tau_{\gamma\gamma}}$, where $\tau_{\gamma\gamma}$ is known as the optical depth, which increases at higher energies. In this case, we apply the model of \cite{Dominguez:2010}. Including this attenuation factor in the expression for the gamma-ray flux (Equation \ref{Int_Flux}), we get:
\begin{equation}
F_{\gamma} (E_{\gamma})=A_{\gamma} \Phi_{\rm input}(\epsilon_{\gamma} )\left (
  \frac{E_{\gamma}}{ \text{TeV}}  \right )^{-\alpha+3} {\rm e}^{-\tau_{\gamma\gamma}(E_{\gamma}, {\rm z})}  \, .
\label{modifiedsed}
\end{equation}

\subsection{Contribution to Mrk 421's Flare}
During the 2010 flaring period of Mrk 421, the VHE data recorded by the IACTs lies in an energy range of 80 GeV to 5 TeV \citep{Aleksic:2014rca}. The energy relation from the Equation \ref{Eegamma} indicates that the seed photon energy in the $p\gamma$ interaction is between 2 and 168 MeV (in the observer's reference frame). The reduction in the source flux combined with the low sensitivity of the LAT below 100 MeV are an impediment to obtaining precise measurements at these energies, and therefore we use the results obtained in Section \ref{sec:fermi_results} to estimate the flux coming from the seed photons ($\Phi_{\rm input} (\epsilon_\gamma)$). The PL input is shown on Figure \ref{Fermi-PL}.

The energy range of the seed photons and the threshold condition for the $\Delta-$resonance can be used to estimate proton energy. From Equation \ref{Eq-7}, we have that $E_{\rm p} \sim 10 E_{\gamma}$ in the observer's reference frame; if measured from Earth, these high energy protons, boosted by the blazar's jet, will be detected in an energy range of $\sim$~800~GeV to 50~TeV, which corresponds in the emission region reference frame to $\sim$ 40~GeV < $E'_{\rm p}$ < 2.45~TeV. This is the energy range of the protons to reach the threshold condition for the $\Delta-$resonance.

\begin{table*}
\centering
\begin{tabular}{cccccccc}
\hline
Time & $A_\gamma$ & $\alpha$ & Preferred & $\Delta {\rm AIC}_{\rm{SSC}, \it{p}\gamma}$ & $\Delta {\rm AIC}_{\rm{two-zone, SSC}, \it{p}\gamma}$ \\
MJD & & & Model & &  \\
\hline
55266& 5.02 $\pm$ 2.74 & 3.12 $\pm$ 0.07 & two-zone SSC & 25.45 & -48.78 \\
55267& 27.24$\pm$ 12.79& 3.41 $\pm$ 0.09 & $p\gamma$ & 6.11 & 9.10 \\
55274& 0.19 $\pm$ 0.01 & 2.31 $\pm$ 0.03 & inconclusive & 2.54 & 0.73 \\
55276& 0.10 $\pm$ 0.02 & 2.17 $\pm$ 0.03 & $p\gamma$ & 26.40 & 2.04 \\
55277& 0.18 $\pm$ 0.02 & 2.32 $\pm$ 0.03 & $p\gamma$ & 5.92 & 2.90 \\
\hline
\end{tabular}
\caption{Summary table of the photohadronic fit for each day in Figure \ref{fits-MJD}. The optimised values for the normalisation constant $A_\gamma$ and the power index $\alpha$ are shown in the second and third column. The AIC difference between the one-zone and two-zone SSC model with respect to the photohadronic ($p\gamma$) model is shown on the fifth and sixth column respectively. An inconclusive result is obtained, if the AIC difference between the two models with the lowest values is less than 2.}
\label{Table-2}
\end{table*}

The emission region has some physical parameters (magnetic field, Doppler factor, radius of the spherical blob) which are the same as those used in the calculation of the photohadronic component. These are fixed parameters taken from the one-zone emission region leptonic model of \cite{Aleksic:2014rca} and are: magnetic field of $B=38$ mG, a Doppler factor ${\cal D}=21$ and a radius of the emission region log$_{10}(R'_{\rm f} [ \text{cm}])=16.72$. The variability reported in \cite{Aleksic:2014rca} corresponds to daily changes in the VHE emission, a time scale which in principle is related to the proton injection. The power index $\alpha$ and the normalisation constant $A_{\gamma}$ were estimated daily for each VHE dataset considered (days at the top of Table \ref{Table-1}). These two free parameters were optimised using a $\chi^2$ minimisation method within \textit{Scipy Python} package \citep{2020SciPy-NMeth}.

\section{Model Fit Results}
\label{sec:model_results}
\begin{figure*}
    \centering
    \resizebox{1.0 \textwidth}{!}{
        \includegraphics{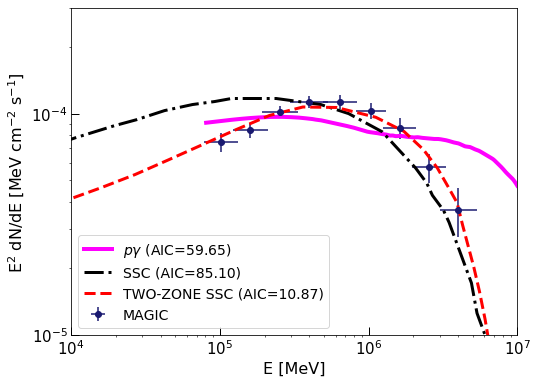}%
        \includegraphics{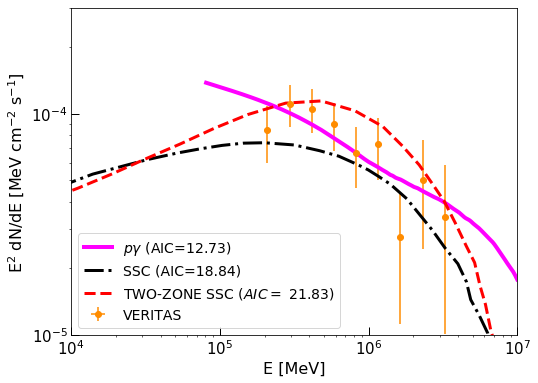}}\\%
        (a) MJD 55266 \hspace{6.5cm} (b) MJD 55267 \\
    \resizebox{1.0 \textwidth}{!}{
        \includegraphics{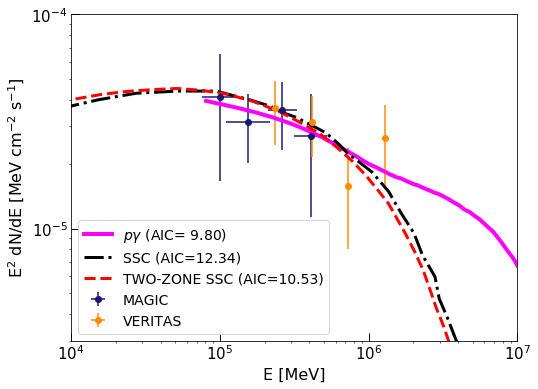}%
        \includegraphics{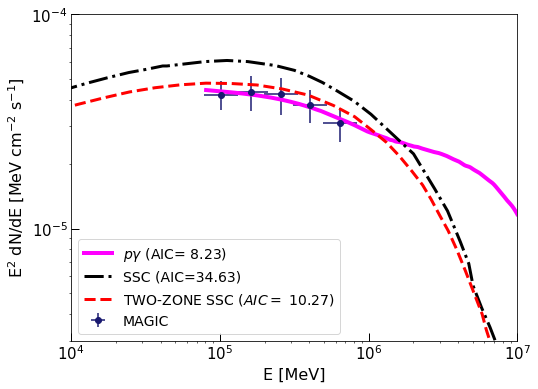}}\\%
        (c)  MJD 55274 \hspace{6.5cm} (d)  MJD 55276 \\
    \resizebox{0.53 \textwidth}{!}{
        \includegraphics{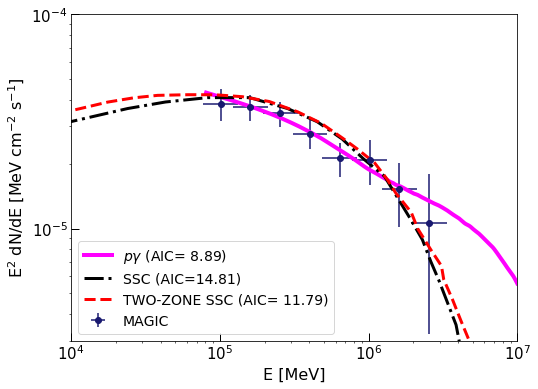}} \\%
        (e)  MJD 55277 \\

    \caption{Photohadronic fit for the VHE gamma-ray data on flaring days with significant TS values: (a) MJD~55266, (b) MJD~55267, (c) MJD~55274, (d) MJD~55276, (e) MJD~55277. The photohadronic component calculated from the PL input is shown in magenta for the valid energy range of the model, which extends roughly down to 80~GeV. The one-zone (two-zone) SSC model from \protect\cite{Aleksic:2014rca} is shown as a dash-dot black (dashed red) line. The calculated AIC values for the three models are included for comparison. VHE data points are from MAGIC and VERITAS observations.}
    \label{fits-MJD}
\end{figure*}

The best-fit value parameters for the photohadronic fit ($A_\gamma$, $\alpha$) are given in Table \ref{Table-2}. The photohadronic component ($p\gamma$ model) for the best-fit values is shown as a magenta continuous curve in Figure~\ref{fits-MJD}; the one-zone leptonic model from \cite{Aleksic:2014rca} (SSC model) is shown as a dash-dot black line, and the two-zone model (two-zone SSC) used in the same paper is represented by a dashed red line. The two-zone SSC model assumes one quiescent blob producing the steady emission, and a smaller independent blob responsible for the temporal evolution of the SED (flaring blob). The VHE energy points in Figure \ref{fits-MJD} are a combination from the IACT observations during the flare (MAGIC and VERITAS, see section~\ref{sec:flare}).

To compare quantitatively these models, we perform an Akaike Information Criterion (AIC) test \citep{Akaike:1973}, which can be used to determine if a model fit is significantly better than another \citep{Harris:2014zqa, Bozdogan:1987, FORSTER:2000}. The AIC takes into account both the goodness of the fit, and the simplicity of the model. This is done by assessing the likelihood and the number of free parameters adopted, AIC is defined as:
\begin{equation}
    {\rm AIC}_{\rm s} = -2 ln(L_{\rm s}) + 2k_{\rm fs} \, ,
\end{equation}
where $L_{\rm s}$ is the likelihood of the model used and $k_{\rm fs}$ is the number of free parameters in the model. In our optimisation process, the $\chi^2$ statistic was used to obtain our best-fit results and is adopted as our likelihood function. Most of the parameters used in the photohadronic model are fixed. There are 4 free parameters considered within the model, which vary to fit the daily VHE observations: 2 from the PL input of the seed photons and 2 from the calculation of the gamma-ray flux $F_\gamma$ from the photohadronic component.

According to \cite{Aleksic:2014rca}, 11 parameters were used in the one-zone leptonic model, 9 of which come from a broken power-law function (with two breaks) in the electron energy distribution required to satisfactorily describe the SED during the flaring period. From these, 5 parameters were left free to vary during the flaring days and the rest were fixed. For the two-zone leptonic model, 20 parameters were used to describe the SED. After fixing the parameters of the quiescent blob, only 4 parameters were left free to vary in the flaring blob. We use the number of free parameters in each model to calculate the corresponding AIC value.

The AIC is based on a theoretical framework within information theory and provides a way to evaluate both the simplicity and accuracy of the model fits. The difference between the AIC of two models p,q is expressed as:
\begin{equation}
    \Delta {\rm AIC}_{\rm p,q} = {\rm AIC}_{\rm p} - {\rm AIC}_{\rm q} \, .
\end{equation}

The AIC difference enables the models considered to be compared and ranked. The model with the lowest AIC represents the best description of the empirical data available. Any model comparison with a $\Delta {\rm AIC}_{\rm p,q} > 2$ above the minimal AIC value is considered significantly worse \citep{Burnham:2001, Lewis&Butler:2011}. 

A direct comparison through the AIC difference test between the $p\gamma$ model and the two leptonic models considered is shown on Table \ref{Table-2}. Among the IACTs observations from the 2010 flaring period from which we were able to fit a spectrum (Figure \ref{fits-MJD}), we found that the photohadronic fit was the preferred model in 3 out of 5 cases according to the AIC comparison test. On one day (MJD~55266), the two-zone SSC model has the minimum AIC by a large difference with respect to the other models. Meanwhile on MJD~55274, the AIC test was inconclusive, as no difference larger than 2 was found between the $p\gamma$ and the two-zone SSC model.

In all cases, the comparison of the $p\gamma$ model with the one SSC model results in a $\Delta {\rm AIC}_{\rm SSC, \it{p}\gamma} >2$, which means the $p\gamma$ model is favoured over the one-zone SSC model and represents a significantly better fit (Table \ref{Table-2}). Nevertheless, this does not necessarily means that the $p\gamma$ model fits all the datasets well.

During the first days of the observations (MJD~55264-67) the source was in its maximum emission state. The gamma-ray flux detected, as shown in Table \ref{Table-1} for the seed photon spectrum reaches a peak on this day, and the spectral index associated is also the highest at $\kappa \sim 2.2$. This behaviour also relates to the optimised free parameters ($\alpha$ and $A_{\gamma}$) which presents their maximum values during these couple of days.

The fitting for MJD~55266 presents a large difference between the calculated AIC values, the two-zone SSC resulting in the lowest AIC and therefore favoured as the best fit for data from this day. The SSC model over-predicts the VHE flux just above 100 GeV, whereas the $p\gamma$ model over-predicts the flux above the TeV energy limit and is flatter below that point, thus underestimating the flux peak. Neither of  these two models represent a good fit to the data, which is reflected in the AIC difference between them and the two-zone SSC model.

On the following day (MJD~55267), using the corresponding PL input to calculate the photohadronic contribution, the overall behaviour of the VERITAS observations can be fitted with the $p\gamma$ model, but the PL behaviour from the input takes over if extended below 100 GeV and predicts an increase in flux; this is a non-physical effect that is outside our validity range and represents a caveat of the model. Nonetheless, the $p\gamma$ model scored the lowest AIC value among the 3 models and the difference in this day is enough to consider it the best fit.

For the final days of the flaring period (MJD~55274 to 55277) three days of VHE data were available, there being no VHE observations on MJD~55275. The PL spectral parameters used as an input for the $p\gamma$ model are very similar for these 3 days. The spectral index during this period is $\kappa \sim 1.1$ and the power index of the proton spectrum is around $\alpha \sim 2.2-2.3$.

During this later part of the flaring period, the lowest AIC values found were from the $p\gamma$ model, indicating these are the best fits to the datasets and ranking the model as the preferred option on MJD~55276 and 55277.

On MJD~55274 the three models predict a similar flux in the 100 GeV to 1 TeV range, but the photohadronic contribution differs at higher energies where it is expected a larger contribution. This is the only day on which the models were tested with data from both MAGIC and VERITAS, although the $p\gamma$ model reached the lowest value from the three, the AIC difference with the two-zone SSC model was not significant enough ($\Delta {\rm AIC}_{\rm SSC, \it{p}\gamma} > 2$) for the $p\gamma$ model to be selected as the preferred model, therefore this day is listed as inconclusive in Table~\ref{Table-2}.

On MJD~55276, the photohadronic model is able to reproduce accurately the VHE gamma-ray data and has the lowest AIC in the period studied. In contrast, the SSC model over-predicts the MAGIC observations, and although the two-zone SSC is a better approximation to the data, the AIC difference in favour of the $p\gamma$ model is enough to select it as the preferred model.

On the final day studied (MJD~55277) both leptonic models are a good approximation to the data; nevertheless, the $p\gamma$ model results in a much lower AIC value due to the accuracy of the fit and its simplicity (in terms of free parameters), remaining as the preferred model. The behaviour of the $p\gamma$ model and the influence of the PL input approximation will be discussed in the next section.

\section{Discussion and outlook}
\label{sec:discussion}
The fits obtained using the $p\gamma$ model over the flaring period are a significantly better description of the VHE data than a purely one-zone leptonic model, due to a combination of the fit quality obtained and the simplicity of the model in terms of the approximations used and the number of free parameters adopted. The two-zone SSC model is more competitive, being selected as the preferred model on MJD~55266 and scoring a similar AIC value to $p\gamma$ model on MJD~55274. It is worth noting that for the AIC calculation we are using an optimistic estimation of the number of free parameters in the leptonic models, following \cite{Aleksic:2014rca} in which only 4 (5) from a total of 20 (11) parameters are considered as free for the two-zone (one-zone) SSC leptonic model during the flaring.

An interesting feature is that the best fits for a photohadronic contribution are obtained in the last 4 days of the flare. In a more complex hadronic scenario, besides the photo-pion production, there is gamma-ray radiation emitted by proton-synchrotron cooling and the synchrotron radiation of the secondary charged particles. These hadronically induced gamma rays are usually in competition with synchrotron and inverse Compton photons radiated by primary electrons considered in the usual leptonic model \citep{Rachen:1998fd}. On some days the hadronic component could be dominant at VHE and then followed by a dominant SSC leptonic component. If the proton injection occurs randomly, there is no preferred time for this to happened during the flare. The interplay between these mechanisms could lead to a time-dependent model with a dominant component at VHE \citep{Diltz:2016aqg,Dimitrakoudis:2012}. The current stage of the model relies on the leptonic contributions to provide the target photon field at MeV energies (presumably the SSC component) and ignores other hadronic components based on the dominant process of photo-pion production through the $\Delta-$resonance. This work aims to be a first step into a more comprehensive lepto-hadronic modelling of flaring blazars.

A common problem with hadronic modelling is the high proton energy required to produce the observed gamma-ray emission. However, in this case the high frequency of the seed photos considered in the $p\gamma$ interactions lowers the energy threshold for the accelerated protons to 40 GeV < $E'_{\rm p}$ < 2.45 TeV in the comoving frame (the emission region), which is below the extreme energies considered in other hadronic models \citep{Mannheim:1992, Muecke:2002bi}. This feature from the $p\gamma$ model can be considered an advantage that would facilitate the conditions for objects like blazars to produce VHE gamma rays and neutrinos from hadronic interactions. Using the Hillas criterion \citep{1984ARA&A..22..425H, Ptitsyna:2008zs, Meli:2007sv} with the considered parameters of the emission region ($B$, ${\cal D}$, $R'_{\rm f}$), a proton could be accelerated up to $E'_{\rm p,max} \sim 650$ TeV at the source, more than sufficient for the $p\gamma$ model to produce the VHE gamma-ray emission.

An interesting feature of the $p\gamma$ model (Figure \ref{fits-MJD}) is the increase of the energy flux above TeV energies. This is where the one-zone SSC model and the $p\gamma$ model differ and needs to be tested in future observations. The forthcoming Cherenkov Telescope Array (CTA) \citep{CTAConsortium:2018tzg} will be critical to differentiating between these scenarios. The expected sensitivity and capabilities of the CTA observatories will enable detection of the gamma-ray flux from Mrk 421 and other near blazars with unprecedented accuracy and instantaneous sensitivity above tens of TeV in energy. 

CTA will also take part in the Neutrino Target of Opportunity (NToO) program to look for gamma-ray counterparts to neutrino alerts in follow-up observations. The large field-of-view (FoV) and the rapidly re-position capabilities of CTA's telescopes working alongside the real time alert program from IceCube \citep{Blaufuss:2019fgv} look very promising for the upcoming years. A more detailed discussion of the current work and development of the NToO for CTA can be found in \cite{NTOO-CTA:2019}.

From the kinematics of Equation \ref{Eq-1}, each $\pi^+$ will produce 3 neutrinos and one $e^+$, which will carry 1/4 of the $\pi^+$ energy each. In the other channel, the $\pi^0$ will produce a pair of photons, so that the observed gamma-ray photon energy and the neutrino energy satisfy $E_\nu \sim E_\gamma / 2$ \citep{Mucke:1998mk}. Also from the kinematics of the decay chain, we have that $F_\nu \sim \frac{3}{4} F_{\pi^+} = \frac{3}{8} F_{\gamma}$ \citep{Sahu:2012wv, Sahu:2013ixa}. After neutrino oscillations, the expected flavor ratio at earth will be $\nu_e : \nu_\mu : \nu_\tau = 1 : 1 : 1$, and so the estimated neutrino flux for only muon neutrinos will be a third of the all flavour flux, then $F_{\nu_\mu} \sim \frac{1}{8} F_{\gamma}$.

For the VHE gamma-ray photons in consideration from these observations, the energy threshold condition for the $\Delta-$resonance leads to a starting point for the neutrino energy range (in the observer's reference frame) of around 0.04 TeV < $E_{\nu, \rm min}$ < 2.5 TeV. From the maximum proton energy condition given by the Hillas criterion, the maximum neutrino energy in the observer's frame is expected to be $E_{\nu, \rm max} \sim 680$ TeV.

If we set an upper limit of $F_{\gamma} < 10^{-4}$ MeV cm$^{-2}$ s$^{-1}$ (which is around the maximum value reached during the flare) and if we assume a neutrino spectrum of the type $\frac{\rm{d} \it{N}}{\rm{d} \it{E}_{\nu}} = A_{\nu} E_{\nu}^{-2}$, then the constant $A_{\nu}$ is estimated as a fraction of the gamma-ray flux. From the highest flux state of the source, we have an approximation of $ A_{\nu} \sim 3.1 \times 10^{-12}$ TeV cm$^{-2}$ s$^{-1}$. Then, by integrating the neutrino spectrum using the effective areas $A_{\rm eff} (E_{\nu})$ of the 59-strings configuration (IC-59) operating in 2010 \citep{Aartsen:2013uuv,Halzen:2005pz}, we can set an upper limit of $ N_{\rm events} < 0.14$ for the expected number neutrino events that would have been detected during the Mrk~421 flaring event. While in this paper, we are focusing on the explanation of the VHE gamma-ray emission rather than neutrino production, the possible detection of neutrino events from flaring blazars is an interesting topic that will be investigated in a future work on the photohadronic contributions.

On MJD~55267, 55274 and 55277, the photohadronic model would behave like a PL that increases, if we extend the fit below $10^5$~MeV. This is related to the type of seed photon input that we are using (a PL approximation) and does not represent a realistic physical description of the SED. This is a caveat of the current model which is focused on the contribution of the photohadronic component at VHE and it is not designed to extend continuously to lower energies. A more complex input model would improve the overall final fit but at the expense of a larger number of free parameters. For the PL description of the seed photon spectrum, 2 free parameters were optimised over bins lasting a couple of days due to the lack of photon statistics; this was the shortest time period over which it was possible to obtain \textit{Fermi} spectra in the >~100~MeV energy range. The parameters $\alpha$ and $A_\gamma$ coming from the photohadronic contribution at VHE were calculated on daily intervals: this discrepancy between the time bins was dictated by the differences in instantaneous sensitivity between the space and ground-based telescopes and might affect the accuracy of the final fit.

The uncertainty of the seed photon spectrum impacts the optimisation process for the free model parameters $\alpha$ and $A_{\gamma}$ which are chosen to get the best-fit values to the VHE data and therefore the minimum AIC value for the model. Better data in the MeV energy range (1 MeV to 100 MeV) would enable a better description of the input seed photons and hence an improvement in the photohadronic fit and more reliable predictions. On this regard, the All-sky Medium Energy Gamma-ray Observatory (AMEGO) mission \citep{AMEGO:2019tcm}, which is planned to operate from 200~keV to >10 ~GeV with $\sim 5 \times$ better angular resolution than \textit{Fermi}-LAT, would be very helpful. The combination of MeV gamma-ray photon data with precise measurements from gamma rays in the 10s of TeV regime would be ideal to further test hadronic emission models.

\section{Conclusions}
\label{sec:conclusions}

 The hadronic modelling of flaring episodes from blazars is a complex challenge that has gained relevance in recent years. In this work, the scenario of a dominant hadronic contribution in the VHE region of the SED coming from $p\gamma$ interactions during the flaring period of Mrk 421 in 2010 was studied. A photohadronic model with 4 free parameters was used and the gamma-ray flux calculated using the $\Delta-$resonance approximation. For the target photon spectrum, we used a PL description estimated from an analysis performed of \textit{Fermi}-LAT data. The injected proton spectrum assumed for the model was also characterised by a PL. We were able to fit the VHE gamma-ray data with the $p\gamma$ model on the days with sufficient photon statistics and according to the AIC test, in all cases the $p\gamma$ model was favoured as a better fit description than a one-zone leptonic model, and in comparison with the two-zone SSC model, the $p\gamma$ was favoured by the AIC test on 3 out of the 5 days fitted (MJD~55267, 55276, 55277). The AIC test was inconclusive on MJD 55274 because the difference between the $p\gamma$ and the two-zone model was not meaningful. On MJD~55266 the two-zone model was favoured as a better description of the observations.

Our results therefore show that a dominant contribution from the photohadronic component can be used to successfully fit the observations of a blazar flaring episode, which shows the potential of including $p\gamma$ interactions in blazar modelling. However, other contributions coming from leptonic processes, synchrotron emission from the charged particles in the hadronic decay chain and cascading effects can also play an important role at VHE. These will be investigated in future works in the search to complement our model towards a lepto-hadronic description.

To explore the neutrino/gamma-ray connection in the upcoming years, the next generation of gamma-ray and neutrino observatories, such as CTA, AMEGO and IceCube-Gen2, will play a crucial role; the improvements in observations at VHE and follow-up programs will make possible to test hadronic components and discriminate between pure leptonic and hadronic scenarios.

\section*{Acknowledgements}
We thank the anonymous referee for their comments and suggestions that helped us to improve the quality and clarity of this article. ARdL acknowledges the support of the National Council for Science and Technology from Mexico (CONACYT). AMB and PMC acknowledge the financial support of the UK Science and Technology Facilities Council consolidated grant ST/P000541/1. 

\section*{Data availability}
This work has made use of public \textit{Fermi} data obtained from the High Energy Astrophysics Science Archive Research Center (HEASARC), provided by NASA Goddard Space Flight Center. The day-by-day broadband SEDs data used from \cite{Aleksic:2014rca} is publicly available at: \url{http://cdsarc.u-strasbg.fr/viz-bin/qcat?J/A+A/578/A22}. This research has made use of \textit{Fermipy} \citep{Wood:Fermipy:2017} to analyse the data from the LAT and \textit{Scipy} (\url{https://www.scipy.org/}) a \textit{Python}-based ecosystem of open-source software for mathematics, science, and engineering. Any extra data generated as part of this project may be shared on a reasonable request to the corresponding author.

\bibliographystyle{mnras}
\bibliography{Bibliography} 

\begin{thebibliography}{}
\makeatletter
\relax
\def\mn@urlcharsother{\let\do\@makeother \do\$\do\&\do\#\do\^\do\_\do\%\do\~}
\def\mn@doi{\begingroup\mn@urlcharsother \@ifnextchar [ {\mn@doi@}
  {\mn@doi@[]}}
\def\mn@doi@[#1]#2{\def\@tempa{#1}\ifx\@tempa\@empty \href
  {http://dx.doi.org/#2} {doi:#2}\else \href {http://dx.doi.org/#2} {#1}\fi
  \endgroup}
\def\mn@eprint#1#2{\mn@eprint@#1:#2::\@nil}
\def\mn@eprint@arXiv#1{\href {http://arxiv.org/abs/#1} {{\tt arXiv:#1}}}
\def\mn@eprint@dblp#1{\href {http://dblp.uni-trier.de/rec/bibtex/#1.xml}
  {dblp:#1}}
\def\mn@eprint@#1:#2:#3:#4\@nil{\def\@tempa {#1}\def\@tempb {#2}\def\@tempc
  {#3}\ifx \@tempc \@empty \let \@tempc \@tempb \let \@tempb \@tempa \fi \ifx
  \@tempb \@empty \def\@tempb {arXiv}\fi \@ifundefined
  {mn@eprint@\@tempb}{\@tempb:\@tempc}{\expandafter \expandafter \csname
  mn@eprint@\@tempb\endcsname \expandafter{\@tempc}}}

\bibitem[\protect\citeauthoryear{{Aartsen} et~al.,}{{Aartsen}
  et~al.}{2013}]{Aartsen:2013uuv}
{Aartsen} M.~G.,  et~al., 2013, \mn@doi [\apj] {10.1088/0004-637X/779/2/132},
  \href {https://ui.adsabs.harvard.edu/abs/2013ApJ...779..132A} {779, 132}

\bibitem[\protect\citeauthoryear{{Abdo} et~al.,}{{Abdo}
  et~al.}{2011}]{Abdo2011ApJ}
{Abdo} A.~A.,  et~al., 2011, \mn@doi [\apj] {10.1088/0004-637X/736/2/131},
  \href {http://adsabs.harvard.edu/abs/2011ApJ...736..131A} {736, 131}

\bibitem[\protect\citeauthoryear{Abdollahi et~al.}{Abdollahi
  et~al.}{2020}]{4FGL:2019}
Abdollahi S.,  et~al., 2020, \mn@doi [\apjs] {10.3847/1538-4365/ab6bcb}, \href
  {https://ui.adsabs.harvard.edu/abs/2020ApJS..247...33A} {247, 33}

\bibitem[\protect\citeauthoryear{{Abeysekara} et~al.,}{{Abeysekara}
  et~al.}{2020}]{Abeysekara:2020}
{Abeysekara} A.~U.,  et~al., 2020, \mn@doi [\apj] {10.3847/1538-4357/ab6612},
  \href {https://ui.adsabs.harvard.edu/abs/2020ApJ...890...97A} {890, 97}

\bibitem[\protect\citeauthoryear{{Aharonian} et~al.,}{{Aharonian}
  et~al.}{2006}]{Aharonian:2005gh}
{Aharonian} F.,  et~al., 2006, \mn@doi [\nat] {10.1038/nature04680}, \href
  {https://ui.adsabs.harvard.edu/abs/2006Natur.440.1018A} {440, 1018}

\bibitem[\protect\citeauthoryear{{Akaike}}{{Akaike}}{1974}]{Akaike:1973}
{Akaike} H.,  1974, IEEE Transactions on Automatic Control, \href
  {https://ui.adsabs.harvard.edu/abs/1974ITAC...19..716A} {19, 716}

\bibitem[\protect\citeauthoryear{{Aleksi{\'c}} et~al.,}{{Aleksi{\'c}}
  et~al.}{2012}]{Aleksic:2012}
{Aleksi{\'c}} J.,  et~al., 2012, \mn@doi [Astroparticle Physics]
  {10.1016/j.astropartphys.2011.11.007}, \href
  {https://ui.adsabs.harvard.edu/abs/2012APh....35..435A} {35, 435}

\bibitem[\protect\citeauthoryear{{Aleksi{\'c}} et~al.,}{{Aleksi{\'c}}
  et~al.}{2015a}]{Aleksic:2015}
{Aleksi{\'c}} J.,  et~al., 2015a, \mn@doi [\aap] {10.1051/0004-6361/201424216},
  \href {https://ui.adsabs.harvard.edu/abs/2015A%26A...576A.126A} {576, A126}

\bibitem[\protect\citeauthoryear{{Aleksi{\'c}} et~al.,}{{Aleksi{\'c}}
  et~al.}{2015b}]{Aleksic:2014rca}
{Aleksi{\'c}} J.,  et~al., 2015b, \mn@doi [\aap] {10.1051/0004-6361/201424811},
  \href {https://ui.adsabs.harvard.edu/abs/2015A&A...578A..22A} {578, A22}

\bibitem[\protect\citeauthoryear{{Araudo}, {Bosch-Ramon}  \& {Romero}}{{Araudo}
  et~al.}{2010}]{Araudo:2010}
{Araudo} A.~T.,  {Bosch-Ramon} V.,   {Romero} G.~E.,  2010, \mn@doi [\aap]
  {10.1051/0004-6361/201014660}, \href
  {https://ui.adsabs.harvard.edu/abs/2010A&A...522A..97A} {522, A97}

\bibitem[\protect\citeauthoryear{{Atwood} et~al.,}{{Atwood}
  et~al.}{2009}]{Atwood:2009}
{Atwood} W.~B.,  et~al., 2009, \mn@doi [\apj] {10.1088/0004-637X/697/2/1071},
  \href {https://ui.adsabs.harvard.edu/abs/2009ApJ...697.1071A} {697, 1071}

\bibitem[\protect\citeauthoryear{{Baring}}{{Baring}}{1997}]{FermiShocks}
{Baring} M.~G.,  1997, in {Giraud-Heraud} Y.,  {Tran Thanh van} J.,  eds, Very
  High Energy Phenomena in the Universe; Moriond Workshop. p.~97 (\mn@eprint
  {arXiv} {astro-ph/9711177})

\bibitem[\protect\citeauthoryear{{Blaufuss}, {Kintscher}, {Lu}  \&
  {Tung}}{{Blaufuss} et~al.}{2019}]{Blaufuss:2019fgv}
{Blaufuss} E.,  {Kintscher} T.,  {Lu} L.,   {Tung} C.~F.,  2019, in 36th
  International Cosmic Ray Conference (ICRC2019). p.~1021 (\mn@eprint {arXiv}
  {1908.04884})

\bibitem[\protect\citeauthoryear{{B{\l}a{\.z}ejowski}
  et~al.,}{{B{\l}a{\.z}ejowski} et~al.}{2005}]{Blaz2005}
{B{\l}a{\.z}ejowski} M.,  et~al., 2005, \mn@doi [\apj] {10.1086/431925}, \href
  {https://ui.adsabs.harvard.edu/abs/2005ApJ...630..130B} {630, 130}

\bibitem[\protect\citeauthoryear{{Bloom} \& {Marscher}}{{Bloom} \&
  {Marscher}}{1996}]{Bloom:1996}
{Bloom} S.~D.,  {Marscher} A.~P.,  1996, \mn@doi [\apj] {10.1086/177092}, \href
  {http://adsabs.harvard.edu/abs/1996ApJ...461..657B} {461, 657}

\bibitem[\protect\citeauthoryear{{B{\"o}ttcher}, {Reimer}, {Sweeney}  \&
  {Prakash}}{{B{\"o}ttcher} et~al.}{2013}]{Boettcher:2013wxa}
{B{\"o}ttcher} M.,  {Reimer} A.,  {Sweeney} K.,   {Prakash} A.,  2013, \mn@doi
  [\apj] {10.1088/0004-637X/768/1/54}, \href
  {https://ui.adsabs.harvard.edu/abs/2013ApJ...768...54B} {768, 54}

\bibitem[\protect\citeauthoryear{Bozdogan}{Bozdogan}{1987}]{Bozdogan:1987}
Bozdogan H.,  1987, \mn@doi [Psychometrika] {10.1007/BF02294361}, 52, 345

\bibitem[\protect\citeauthoryear{{Brown}, {Adams}  \& {Chadwick}}{{Brown}
  et~al.}{2015}]{Brown:2015zxa}
{Brown} A.~M.,  {Adams} J.,   {Chadwick} P.~M.,  2015, \mn@doi [\mnras]
  {10.1093/mnras/stv899}, \href
  {https://ui.adsabs.harvard.edu/abs/2015MNRAS.451..323B} {451, 323}

\bibitem[\protect\citeauthoryear{Burnham \& Anderson}{Burnham \&
  Anderson}{2001}]{Burnham:2001}
Burnham K.~P.,  Anderson D.~R.,  2001, \mn@doi [Wildlife Research]
  {10.1071/wr99107}, 28, 111

\bibitem[\protect\citeauthoryear{{CTA Consortium} et~al.,}{{CTA Consortium}
  et~al.}{2019}]{CTAConsortium:2018tzg}
{CTA Consortium} et~al., 2019, {Science with the Cherenkov Telescope Array}.
WSP, \mn@doi{10.1142/10986}

\bibitem[\protect\citeauthoryear{{Cerruti}, {Zech}, {Boisson}, {Emery}, {Inoue}
   \& {Lenain}}{{Cerruti} et~al.}{2019}]{Cerruti:2018tmc}
{Cerruti} M.,  {Zech} A.,  {Boisson} C.,  {Emery} G.,  {Inoue} S.,   {Lenain}
  J.~P.,  2019, \mn@doi [\mnras] {10.1093/mnrasl/sly210}, \href
  {https://ui.adsabs.harvard.edu/abs/2019MNRAS.483L..12C} {483, L12}

\bibitem[\protect\citeauthoryear{{Dar} \& {Laor}}{{Dar} \&
  {Laor}}{1997}]{Dar:1996qv}
{Dar} A.,  {Laor} A.,  1997, \mn@doi [\apjl] {10.1086/310544}, \href
  {https://ui.adsabs.harvard.edu/abs/1997ApJ...478L...5D} {478, L5}

\bibitem[\protect\citeauthoryear{{Dermer} \& {Schlickeiser}}{{Dermer} \&
  {Schlickeiser}}{1993}]{Dermer:1993}
{Dermer} C.~D.,  {Schlickeiser} R.,  1993, \mn@doi [\apj] {10.1086/173251},
  \href {https://ui.adsabs.harvard.edu/abs/1993ApJ...416..458D} {416, 458}

\bibitem[\protect\citeauthoryear{{Diltz} \& {B{\"o}ttcher}}{{Diltz} \&
  {B{\"o}ttcher}}{2016}]{Diltz:2016aqg}
{Diltz} C.,  {B{\"o}ttcher} M.,  2016, \mn@doi [\apj]
  {10.3847/0004-637X/826/1/54}, \href
  {https://ui.adsabs.harvard.edu/abs/2016ApJ...826...54D} {826, 54}

\bibitem[\protect\citeauthoryear{{Diltz}, {B{\"o}ttcher}  \& {Fossati}}{{Diltz}
  et~al.}{2015}]{Diltz:2015kha}
{Diltz} C.,  {B{\"o}ttcher} M.,   {Fossati} G.,  2015, \mn@doi [\apj]
  {10.1088/0004-637X/802/2/133}, \href
  {https://ui.adsabs.harvard.edu/abs/2015ApJ...802..133D} {802, 133}

\bibitem[\protect\citeauthoryear{{Dimitrakoudis}, {Mastichiadis}, {Protheroe}
  \& {Reimer}}{{Dimitrakoudis} et~al.}{2012}]{Dimitrakoudis:2012}
{Dimitrakoudis} S.,  {Mastichiadis} A.,  {Protheroe} R.~J.,   {Reimer} A.,
  2012, \mn@doi [\aap] {10.1051/0004-6361/201219770}, \href
  {https://ui.adsabs.harvard.edu/abs/2012A&A...546A.120D} {546, A120}

\bibitem[\protect\citeauthoryear{{Dom{\'\i}nguez} et~al.,}{{Dom{\'\i}nguez}
  et~al.}{2011}]{Dominguez:2010}
{Dom{\'\i}nguez} A.,  et~al., 2011, \mn@doi [\mnras]
  {10.1111/j.1365-2966.2010.17631.x}, \href
  {https://ui.adsabs.harvard.edu/abs/2011MNRAS.410.2556D} {410, 2556}

\bibitem[\protect\citeauthoryear{Finke}{Finke}{2018}]{Finke:2016vts}
Finke J.~D.,  2018, \mn@doi [\apj] {10.3847/1538-4357/aac9c6}, \href
  {https://ui.adsabs.harvard.edu/abs/2018ApJ...860..178F} {860, 178}

\bibitem[\protect\citeauthoryear{Forster}{Forster}{2000}]{FORSTER:2000}
Forster M.~R.,  2000, \mn@doi [Journal of Mathematical Psychology]
  {https://doi.org/10.1006/jmps.1999.1284}, 44, 205

\bibitem[\protect\citeauthoryear{Gaisser, Halzen  \& Stanev}{Gaisser
  et~al.}{1995}]{Gaisser:1994yf}
Gaisser T.~K.,  Halzen F.,   Stanev T.,  1995, \mn@doi [Physics Reports]
  {https://doi.org/10.1016/0370-1573(95)00003-Y}, 258, 173

\bibitem[\protect\citeauthoryear{{Ghisellini}, {Tavecchio}  \&
  {Chiaberge}}{{Ghisellini} et~al.}{2005}]{Ghisellini:2005}
{Ghisellini} G.,  {Tavecchio} F.,   {Chiaberge} M.,  2005, \mn@doi [\aap]
  {10.1051/0004-6361:20041404}, \href
  {https://ui.adsabs.harvard.edu/abs/2005A&A...432..401G} {432, 401}

\bibitem[\protect\citeauthoryear{{Giommi}, {Padovani}, {Oikonomou}, {Glauch},
  {Paiano}  \& {Resconi}}{{Giommi} et~al.}{2020}]{2020arXiv200306405G}
{Giommi} P.,  {Padovani} P.,  {Oikonomou} F.,  {Glauch} T.,  {Paiano} S.,
  {Resconi} E.,  2020, arXiv e-prints, \href
  {https://ui.adsabs.harvard.edu/abs/2020arXiv200306405G} {p. arXiv:2003.06405}

\bibitem[\protect\citeauthoryear{{Gl{\"u}senkamp}}{{Gl{\"u}senkamp}}{2016}]{Glusenkamp:2015jca}
{Gl{\"u}senkamp} T.,  2016, in European Physical Journal Web of Conferences. p.
  05006 (\mn@eprint {arXiv} {1502.03104}),
  \mn@doi{10.1051/epjconf/201612105006}

\bibitem[\protect\citeauthoryear{{Halzen} \& {Hooper}}{{Halzen} \&
  {Hooper}}{2005}]{Halzen:2005pz}
{Halzen} F.,  {Hooper} D.,  2005, \mn@doi [Astroparticle Physics]
  {10.1016/j.astropartphys.2005.03.007}, \href
  {https://ui.adsabs.harvard.edu/abs/2005APh....23..537H} {23, 537}

\bibitem[\protect\citeauthoryear{{Halzen}, {Kheirandish}, {Weisgarber}  \&
  {Wakely}}{{Halzen} et~al.}{2019}]{Halzen:2018iak}
{Halzen} F.,  {Kheirandish} A.,  {Weisgarber} T.,   {Wakely} S.~P.,  2019,
  \mn@doi [\apjl] {10.3847/2041-8213/ab0d27}, \href
  {https://ui.adsabs.harvard.edu/abs/2019ApJ...874L...9H} {874, L9}

\bibitem[\protect\citeauthoryear{{Harris}, {Chadwick}  \& {Daniel}}{{Harris}
  et~al.}{2014}]{Harris:2014zqa}
{Harris} J.,  {Chadwick} P.~M.,   {Daniel} M.~K.,  2014, \mn@doi [\mnras]
  {10.1093/mnras/stu787}, \href
  {https://ui.adsabs.harvard.edu/abs/2014MNRAS.441.3591H} {441, 3591}

\bibitem[\protect\citeauthoryear{{Hillas}}{{Hillas}}{1984}]{1984ARA&A..22..425H}
{Hillas} A.~M.,  1984, \mn@doi [\araa] {10.1146/annurev.aa.22.090184.002233},
  \href {https://ui.adsabs.harvard.edu/abs/1984ARA&A..22..425H} {22, 425}

\bibitem[\protect\citeauthoryear{{Homan}, {Lister}, {Kovalev}, {Pushkarev},
  {Savolainen}, {Kellermann}, {Richards}  \& {Ros}}{{Homan}
  et~al.}{2015}]{MOJAVE:2015}
{Homan} D.~C.,  {Lister} M.~L.,  {Kovalev} Y.~Y.,  {Pushkarev} A.~B.,
  {Savolainen} T.,  {Kellermann} K.~I.,  {Richards} J.~L.,   {Ros} E.,  2015,
  \mn@doi [\apj] {10.1088/0004-637X/798/2/134}, \href
  {https://ui.adsabs.harvard.edu/abs/2015ApJ...798..134H} {798, 134}

\bibitem[\protect\citeauthoryear{{H{\"u}mmer}, {R{\"u}ger}, {Spanier}  \&
  {Winter}}{{H{\"u}mmer} et~al.}{2010}]{Hummer:2010vx}
{H{\"u}mmer} S.,  {R{\"u}ger} M.,  {Spanier} F.,   {Winter} W.,  2010, \mn@doi
  [\apj] {10.1088/0004-637X/721/1/630}, \href
  {https://ui.adsabs.harvard.edu/abs/2010ApJ...721..630H} {721, 630}

\bibitem[\protect\citeauthoryear{{IceCube Collaboration} et~al.,}{{IceCube
  Collaboration} et~al.}{2018a}]{IceCube:2018cha}
{IceCube Collaboration} et~al., 2018a, \mn@doi [Science]
  {10.1126/science.aat2890}, \href
  {https://ui.adsabs.harvard.edu/abs/2018Sci...361..147I} {361, 147}

\bibitem[\protect\citeauthoryear{{IceCube Collaboration} et~al.,}{{IceCube
  Collaboration} et~al.}{2018b}]{MultiW:2018dnn}
{IceCube Collaboration} et~al., 2018b, \mn@doi [Science]
  {10.1126/science.aat1378}, \href
  {https://ui.adsabs.harvard.edu/abs/2018Sci...361.1378I} {361, eaat1378}

\bibitem[\protect\citeauthoryear{{Kadler} et~al.,}{{Kadler}
  et~al.}{2016}]{Kadler:2016ygj}
{Kadler} M.,  et~al., 2016, \mn@doi [Nature Physics] {10.1038/nphys3715}, \href
  {https://ui.adsabs.harvard.edu/abs/2016NatPh..12..807K} {12, 807}

\bibitem[\protect\citeauthoryear{{Kildea} et~al.,}{{Kildea}
  et~al.}{2007}]{Kildea2007}
{Kildea} J.,  et~al., 2007, \mn@doi [Astroparticle Physics]
  {10.1016/j.astropartphys.2007.05.004}, \href
  {https://ui.adsabs.harvard.edu/abs/2007APh....28..182K} {28, 182}

\bibitem[\protect\citeauthoryear{{Krau{\ss}} et~al.,}{{Krau{\ss}}
  et~al.}{2014}]{Krauss:2014tna}
{Krau{\ss}} F.,  et~al., 2014, \mn@doi [\aap] {10.1051/0004-6361/201424219},
  \href {https://ui.adsabs.harvard.edu/abs/2014A&A...566L...7K} {566, L7}

\bibitem[\protect\citeauthoryear{Krau{\ss} et~al.}{Krau{\ss}
  et~al.}{2015}]{Krauss:2015pja}
Krau{\ss} F.,  et~al., 2015, in {5th International Fermi Symposium}.
  (\mn@eprint {arXiv} {1502.02147})

\bibitem[\protect\citeauthoryear{Lewis, Butler  \& Gilbert}{Lewis
  et~al.}{2011}]{Lewis&Butler:2011}
Lewis F.,  Butler A.,   Gilbert L.,  2011, \mn@doi [Methods in Ecology and
  Evolution] {10.1111/j.2041-210X.2010.00063.x}, 2, 155

\bibitem[\protect\citeauthoryear{{Liu}, {Wang}, {Xue}, {Taylor}, {Wang}, {Li}
  \& {Yan}}{{Liu} et~al.}{2019}]{Ruo:2019}
{Liu} R.-Y.,  {Wang} K.,  {Xue} R.,  {Taylor} A.~M.,  {Wang} X.-Y.,  {Li} Z.,
  {Yan} H.,  2019, \mn@doi [\prd] {10.1103/PhysRevD.99.063008}, \href
  {https://ui.adsabs.harvard.edu/abs/2019PhRvD..99f3008L} {99, 063008}

\bibitem[\protect\citeauthoryear{{MacDonald}, {Marscher}, {Jorstad}  \&
  {Joshi}}{{MacDonald} et~al.}{2015}]{MacDonald:2015}
{MacDonald} N.~R.,  {Marscher} A.~P.,  {Jorstad} S.~G.,   {Joshi} M.,  2015,
  \mn@doi [\apj] {10.1088/0004-637X/804/2/111}, \href
  {https://ui.adsabs.harvard.edu/abs/2015ApJ...804..111M} {804, 111}

\bibitem[\protect\citeauthoryear{{Mannheim}}{{Mannheim}}{1993}]{Mannheim:1993}
{Mannheim} K.,  1993, \aap, \href
  {https://ui.adsabs.harvard.edu/abs/1993A&A...269...67M} {269, 67}

\bibitem[\protect\citeauthoryear{{Mannheim} \& {Biermann}}{{Mannheim} \&
  {Biermann}}{1992}]{Mannheim:1992}
{Mannheim} K.,  {Biermann} P.~L.,  1992, \aap, \href
  {https://ui.adsabs.harvard.edu/abs/1992A&A...253L..21M} {253, L21}

\bibitem[\protect\citeauthoryear{{Maraschi}, {Ghisellini}  \&
  {Celotti}}{{Maraschi} et~al.}{1992}]{Maraschi:1993}
{Maraschi} L.,  {Ghisellini} G.,   {Celotti} A.,  1992, \mn@doi [\apjl]
  {10.1086/186531}, \href
  {https://ui.adsabs.harvard.edu/abs/1992ApJ...397L...5M} {397, L5}

\bibitem[\protect\citeauthoryear{{Marscher} et~al.,}{{Marscher}
  et~al.}{2008}]{Marscher:2008}
{Marscher} A.~P.,  et~al., 2008, \mn@doi [\nat] {10.1038/nature06895}, \href
  {https://ui.adsabs.harvard.edu/abs/2008Natur.452..966M} {452, 966}

\bibitem[\protect\citeauthoryear{{Marscher} et~al.,}{{Marscher}
  et~al.}{2010}]{Marscher:2010}
{Marscher} A.~P.,  et~al., 2010, \mn@doi [\apjl]
  {10.1088/2041-8205/710/2/L126}, \href
  {https://ui.adsabs.harvard.edu/abs/2010ApJ...710L.126M} {710, L126}

\bibitem[\protect\citeauthoryear{{Mattox} et~al.,}{{Mattox}
  et~al.}{1996}]{Mattox:1996}
{Mattox} J.~R.,  et~al., 1996, \mn@doi [\apj] {10.1086/177068}, 461, 396

\bibitem[\protect\citeauthoryear{{McEnery} et~al.,}{{McEnery}
  et~al.}{2019}]{AMEGO:2019tcm}
{McEnery} J.,  et~al., 2019, in \baas. p.~245 (\mn@eprint {arXiv} {1907.07558})

\bibitem[\protect\citeauthoryear{{Meli}, {Becker}  \& {Quenby}}{{Meli}
  et~al.}{2008}]{Meli:2007sv}
{Meli} A.,  {Becker} J.~K.,   {Quenby} J.~J.,  2008, \mn@doi [\aap]
  {10.1051/0004-6361:20078681}, \href
  {https://ui.adsabs.harvard.edu/abs/2008A&A...492..323M} {492, 323}

\bibitem[\protect\citeauthoryear{{M{\"u}cke} \& {Protheroe}}{{M{\"u}cke} \&
  {Protheroe}}{2001}]{Mucke:2000rn}
{M{\"u}cke} A.,  {Protheroe} R.~J.,  2001, \mn@doi [Astroparticle Physics]
  {10.1016/S0927-6505(00)00141-9}, \href
  {https://ui.adsabs.harvard.edu/abs/2001APh....15..121M} {15, 121}

\bibitem[\protect\citeauthoryear{{M{\"u}cke}, {Rachen}, {Engel}, {Protheroe}
  \& {Stanev}}{{M{\"u}cke} et~al.}{1999}]{Mucke:1998mk}
{M{\"u}cke} A.,  {Rachen} J.~P.,  {Engel} R.,  {Protheroe} R.~J.,   {Stanev}
  T.,  1999, \mn@doi [\pasa] {10.1071/AS99160}, \href
  {https://ui.adsabs.harvard.edu/abs/1999PASA...16..160M} {16, 160}

\bibitem[\protect\citeauthoryear{{M{\"u}cke}, {Protheroe}, {Engel}, {Rachen}
  \& {Stanev}}{{M{\"u}cke} et~al.}{2003}]{Muecke:2002bi}
{M{\"u}cke} A.,  {Protheroe} R.~J.,  {Engel} R.,  {Rachen} J.~P.,   {Stanev}
  T.,  2003, \mn@doi [Astroparticle Physics] {10.1016/S0927-6505(02)00185-8},
  \href {https://ui.adsabs.harvard.edu/abs/2003APh....18..593M} {18, 593}

\bibitem[\protect\citeauthoryear{{Murase}, {Oikonomou}  \&
  {Petropoulou}}{{Murase} et~al.}{2018}]{Murase:2018iyl}
{Murase} K.,  {Oikonomou} F.,   {Petropoulou} M.,  2018, \mn@doi [\apj]
  {10.3847/1538-4357/aada00}, \href
  {https://ui.adsabs.harvard.edu/abs/2018ApJ...865..124M} {865, 124}

\bibitem[\protect\citeauthoryear{{Oikonomou}, {Murase}, {Padovani}, {Resconi}
  \& {M{\'e}sz{\'a}ros}}{{Oikonomou} et~al.}{2019}]{2019:Oikonomou}
{Oikonomou} F.,  {Murase} K.,  {Padovani} P.,  {Resconi} E.,
  {M{\'e}sz{\'a}ros} P.,  2019, \mn@doi [\mnras] {10.1093/mnras/stz2246}, \href
  {https://ui.adsabs.harvard.edu/abs/2019MNRAS.489.4347O} {489, 4347}

\bibitem[\protect\citeauthoryear{{Padovani}, {Resconi}, {Giommi}, {Arsioli}  \&
  {Chang}}{{Padovani} et~al.}{2016}]{Padovani:2016wwn}
{Padovani} P.,  {Resconi} E.,  {Giommi} P.,  {Arsioli} B.,   {Chang} Y.~L.,
  2016, \mn@doi [\mnras] {10.1093/mnras/stw228}, \href
  {https://ui.adsabs.harvard.edu/abs/2016MNRAS.457.3582P} {457, 3582}

\bibitem[\protect\citeauthoryear{{Padovani}, {Oikonomou}, {Petropoulou},
  {Giommi}  \& {Resconi}}{{Padovani} et~al.}{2019}]{Padovani:2019xcv}
{Padovani} P.,  {Oikonomou} F.,  {Petropoulou} M.,  {Giommi} P.,   {Resconi}
  E.,  2019, \mn@doi [\mnras] {10.1093/mnrasl/slz011}, \href
  {https://ui.adsabs.harvard.edu/abs/2019MNRAS.484L.104P} {484, L104}

\bibitem[\protect\citeauthoryear{{Palladino}, {Rodrigues}, {Gao}  \&
  {Winter}}{{Palladino} et~al.}{2019}]{Palladino:2019}
{Palladino} A.,  {Rodrigues} X.,  {Gao} S.,   {Winter} W.,  2019, \mn@doi
  [\apj] {10.3847/1538-4357/aaf507}, \href
  {https://ui.adsabs.harvard.edu/abs/2019ApJ...871...41P} {871, 41}

\bibitem[\protect\citeauthoryear{{Perkins}, {Maier}  \& {The VERITAS
  Collaboration}}{{Perkins} et~al.}{2009}]{Perkins2009}
{Perkins} J.~S.,  {Maier} G.,   {The VERITAS Collaboration} 2009, arXiv
  e-prints, \href {https://ui.adsabs.harvard.edu/abs/2009arXiv0912.3841P} {p.
  arXiv:0912.3841}

\bibitem[\protect\citeauthoryear{{Petropoulou} et~al.,}{{Petropoulou}
  et~al.}{2020}]{Petropoulou:2019zqp}
{Petropoulou} M.,  et~al., 2020, \mn@doi [\apj] {10.3847/1538-4357/ab76d0},
  \href {https://ui.adsabs.harvard.edu/abs/2020ApJ...891..115P} {891, 115}

\bibitem[\protect\citeauthoryear{{Pichel}}{{Pichel}}{2009}]{Pichel:2009}
{Pichel} A.,  2009, arXiv e-prints, \href
  {https://ui.adsabs.harvard.edu/abs/2009arXiv0908.0010P} {p. arXiv:0908.0010}

\bibitem[\protect\citeauthoryear{{Ptitsyna} \& {Troitsky}}{{Ptitsyna} \&
  {Troitsky}}{2010}]{Ptitsyna:2008zs}
{Ptitsyna} K.~V.,  {Troitsky} S.~V.,  2010, \mn@doi [Physics Uspekhi]
  {10.3367/UFNe.0180.201007c.0723}, \href
  {https://ui.adsabs.harvard.edu/abs/2010PhyU...53..691P} {53, 691}

\bibitem[\protect\citeauthoryear{{Punch} et~al.,}{{Punch}
  et~al.}{1992}]{Punch:1992}
{Punch} M.,  et~al., 1992, \mn@doi [\nat] {10.1038/358477a0}, \href
  {https://ui.adsabs.harvard.edu/abs/1992Natur.358..477P} {358, 477}

\bibitem[\protect\citeauthoryear{{Rachen} \& {M{\'e}sz{\'a}ros}}{{Rachen} \&
  {M{\'e}sz{\'a}ros}}{1998}]{Rachen:1998fd}
{Rachen} J.~P.,  {M{\'e}sz{\'a}ros} P.,  1998, \mn@doi [\prd]
  {10.1103/PhysRevD.58.123005}, \href
  {https://ui.adsabs.harvard.edu/abs/1998PhRvD..58l3005R} {58, 123005}

\bibitem[\protect\citeauthoryear{{Righi}, {Tavecchio}  \& {Inoue}}{{Righi}
  et~al.}{2019}]{Righi:2018}
{Righi} C.,  {Tavecchio} F.,   {Inoue} S.,  2019, \mn@doi [\mnras]
  {10.1093/mnrasl/sly231}, \href
  {https://ui.adsabs.harvard.edu/abs/2019MNRAS.483L.127R} {483, L127}

\bibitem[\protect\citeauthoryear{{Rodrigues}, {Gao}, {Fedynitch}, {Palladino}
  \& {Winter}}{{Rodrigues} et~al.}{2019}]{Rodrigues:2018tku}
{Rodrigues} X.,  {Gao} S.,  {Fedynitch} A.,  {Palladino} A.,   {Winter} W.,
  2019, \mn@doi [\apjl] {10.3847/2041-8213/ab1267}, \href
  {https://ui.adsabs.harvard.edu/abs/2019ApJ...874L..29R} {874, L29}

\bibitem[\protect\citeauthoryear{{Sahayanathan} \& {Godambe}}{{Sahayanathan} \&
  {Godambe}}{2012}]{Sahayan:2011}
{Sahayanathan} S.,  {Godambe} S.,  2012, \mn@doi [\mnras]
  {10.1111/j.1365-2966.2011.19829.x}, \href
  {https://ui.adsabs.harvard.edu/abs/2012MNRAS.419.1660S} {419, 1660}

\bibitem[\protect\citeauthoryear{{Sahu}, {Zhang}  \& {Fraija}}{{Sahu}
  et~al.}{2012}]{Sahu:2012wv}
{Sahu} S.,  {Zhang} B.,   {Fraija} N.,  2012, \mn@doi [\prd]
  {10.1103/PhysRevD.85.043012}, \href
  {https://ui.adsabs.harvard.edu/abs/2012PhRvD..85d3012S} {85, 043012}

\bibitem[\protect\citeauthoryear{{Sahu}, {Oliveros}  \& {Sanabria}}{{Sahu}
  et~al.}{2013}]{Sahu:2013ixa}
{Sahu} S.,  {Oliveros} A. F.~O.,   {Sanabria} J.~C.,  2013, \mn@doi [\prd]
  {10.1103/PhysRevD.87.103015}, \href
  {https://ui.adsabs.harvard.edu/abs/2013PhRvD..87j3015S} {87, 103015}

\bibitem[\protect\citeauthoryear{{Sahu}, {Miranda}  \& {Rajpoot}}{{Sahu}
  et~al.}{2016}]{Sahu:2015tua}
{Sahu} S.,  {Miranda} L.~S.,   {Rajpoot} S.,  2016, \mn@doi [Eur. Phys. J. C]
  {10.1140/epjc/s10052-016-3975-2}, \href
  {https://ui.adsabs.harvard.edu/abs/2016EPJC...76..127S} {76, 127}

\bibitem[\protect\citeauthoryear{{Sahu}, {de Le{\'o}n}  \& {Miranda}}{{Sahu}
  et~al.}{2017}]{Sahu:2016bdu}
{Sahu} S.,  {de Le{\'o}n} A.~R.,   {Miranda} L.~S.,  2017, \mn@doi [Eur. Phys.
  J. C] {10.1140/epjc/s10052-017-5335-2}, \href
  {https://ui.adsabs.harvard.edu/abs/2017EPJC...77..741S} {77, 741}

\bibitem[\protect\citeauthoryear{{Sahu}, {de Le{\'o}n}  \& {Nagataki}}{{Sahu}
  et~al.}{2018a}]{Sahu:2018rqs}
{Sahu} S.,  {de Le{\'o}n} A.~R.,   {Nagataki} S.,  2018a, \mn@doi [Eur. Phys.
  J. C] {10.1140/epjc/s10052-018-5954-2}, \href
  {https://ui.adsabs.harvard.edu/abs/2018EPJC...78..484S} {78, 484}

\bibitem[\protect\citeauthoryear{{Sahu}, {de Le{\'o}n}, {Nagataki}  \&
  {Gupta}}{{Sahu} et~al.}{2018b}]{Sahu:2018gik}
{Sahu} S.,  {de Le{\'o}n} A.~R.,  {Nagataki} S.,   {Gupta} V.,  2018b, \mn@doi
  [Eur. Phys. J. C] {10.1140/epjc/s10052-018-6038-z}, \href
  {https://ui.adsabs.harvard.edu/abs/2018EPJC...78..557S} {78, 557}

\bibitem[\protect\citeauthoryear{Satalecka et~al.,}{Satalecka
  et~al.}{2020}]{NTOO-CTA:2019}
Satalecka K.,  et~al., 2020, \mn@doi [PoS] {10.22323/1.358.0784}, ICRC2019, 784

\bibitem[\protect\citeauthoryear{{Sikora}, {Begelman}  \& {Rees}}{{Sikora}
  et~al.}{1994}]{Sikora:1994}
{Sikora} M.,  {Begelman} M.~C.,   {Rees} M.~J.,  1994, \mn@doi [\apj]
  {10.1086/173633}, \href
  {https://ui.adsabs.harvard.edu/abs/1994ApJ...421..153S} {421, 153}

\bibitem[\protect\citeauthoryear{{Tanaka}, {Buson}  \& {Kocevski}}{{Tanaka}
  et~al.}{2017}]{MMS-Tanaka-Tel}
{Tanaka} Y.~T.,  {Buson} S.,   {Kocevski} D.,  2017, The Astronomer's Telegram,
  \href {https://ui.adsabs.harvard.edu/abs/2017ATel10791....1T} {10791, 1}

\bibitem[\protect\citeauthoryear{{Tavecchio}, {Maraschi}  \&
  {Ghisellini}}{{Tavecchio} et~al.}{1998}]{Tavecchio:1998}
{Tavecchio} F.,  {Maraschi} L.,   {Ghisellini} G.,  1998, \mn@doi [\apj]
  {10.1086/306526}, \href
  {https://ui.adsabs.harvard.edu/abs/1998ApJ...509..608T} {509, 608}

\bibitem[\protect\citeauthoryear{{Tavecchio}, {Ghisellini}, {Ghirlanda},
  {Costamante}  \& {Franceschini}}{{Tavecchio} et~al.}{2009}]{Tavecchio:2009}
{Tavecchio} F.,  {Ghisellini} G.,  {Ghirlanda} G.,  {Costamante} L.,
  {Franceschini} A.,  2009, \mn@doi [\mnras]
  {10.1111/j.1745-3933.2009.00724.x}, \href
  {https://ui.adsabs.harvard.edu/abs/2009MNRAS.399L..59T} {399, L59}

\bibitem[\protect\citeauthoryear{{Virtanen} et~al.,}{{Virtanen}
  et~al.}{2020}]{2020SciPy-NMeth}
{Virtanen} P.,  et~al., 2020, \mn@doi [Nature Methods]
  {10.1038/s41592-019-0686-2}, \href
  {https://ui.adsabs.harvard.edu/abs/2020NatMe..17..261V} {17, 261}

\bibitem[\protect\citeauthoryear{{Walker}, {Hardee}, {Davies}, {Ly}, {Junor},
  {Mertens}  \& {Lobanov}}{{Walker} et~al.}{2016}]{Walker:2016}
{Walker} R.~C.,  {Hardee} P.~E.,  {Davies} F.,  {Ly} C.,  {Junor} W.,
  {Mertens} F.,   {Lobanov} A.,  2016, \mn@doi [Galaxies]
  {10.3390/galaxies4040046}, \href
  {https://ui.adsabs.harvard.edu/abs/2016Galax...4...46W} {4, 46}

\bibitem[\protect\citeauthoryear{{Winter} \& {Gao}}{{Winter} \&
  {Gao}}{2019}]{Winter:2019hee}
{Winter} W.,  {Gao} S.,  2019, in 36th International Cosmic Ray Conference
  (ICRC2019). p.~1032 (\mn@eprint {arXiv} {1909.06289})

\bibitem[\protect\citeauthoryear{{Wood}, {Caputo}, {Charles}, {Di Mauro},
  {Magill}, {Perkins}  \& {Fermi-LAT Collaboration}}{{Wood}
  et~al.}{2017}]{Wood:Fermipy:2017}
{Wood} M.,  {Caputo} R.,  {Charles} E.,  {Di Mauro} M.,  {Magill} J.,
  {Perkins} J.~S.,   {Fermi-LAT Collaboration} 2017, in 35th International
  Cosmic Ray Conference (ICRC2017). p.~824 (\mn@eprint {arXiv} {1707.09551})

\makeatother
\end{thebibliography}

\bsp	
\label{lastpage}
\end{document}